\documentclass{aastex631}

\usepackage{multirow}
\usepackage{booktabs}
\usepackage{amsmath}
\usepackage{subfigure}

\begin{document}

\title{Dynamic Three-dimensional Simulation of Surface Charging on Rotating Asteroids}

\author[0000-0002-4321-9008]{Ronghui Quan}
\affiliation{College of Astronautics, Nanjing University of Aeronautics and Astronautics, Nanjing, Jiangsu, 210016, China}

\author[0000-0003-3498-2172]{Zhiying Song}
\altaffiliation{Correspondence to: Zhiying Song, sx2215054@nuaa.edu.cn}
\affiliation{College of Astronautics, Nanjing University of Aeronautics and Astronautics, Nanjing, Jiangsu, 210016, China}

\author[0000-0002-9859-0277]{Zhigui Liu}
\affiliation{College of Astronautics, Nanjing University of Aeronautics and Astronautics, Nanjing, Jiangsu, 210016, China}

\begin{abstract}

Surface charging phenomenon of asteroids, mainly resulting from solar wind plasma and solar radiation, has been studied extensively. However, the influence of asteroid's rotation on surface charging has yet to be fully understood. Here neural network is established to replace numerical integration, improving the efficiency of dynamic three-dimensional simulation. We implement simulation of rotating asteroids and surrounding plasma environment under different conditions, including quiet solar wind and solar storms, various minerals on asteroid's surface also be considered. For asteroids with rotation periods comparable to orbital period, effect of orbital motion and obliquity also be studied. Results show that under typical solar wind, the maximum and minimum potential of asteroids will gradually decrease with their increasing periods, especially when solar wind is obliquely incident. For asteroid has period longer than one week, this decreasing trend will become extremely slow. During solar storm passing, solar wind plasma changes sharply, the susceptibility of asteroid's surface potential to rotation is greatly pronounced. Minerals on surface also count, plagioclase is the most sensitive mineral among those we explored, while ilmenite seems indifferent to changes in rotation periods. Understanding the surface charging of asteroid under various rotation periods or angles, is crucial for further research into solar wind plasma and asteroid's surface dust motion, providing a reference for safe landing exploration of asteroids.

\end{abstract}


\section{Introduction} \label{sec:intro}

Small airless bodies such as comets and asteroids exposed to solar wind plasma and ultraviolet irradiation, are very sensitive to changes in surrounding magnetic and electric fields. Their exhibit surface charging phenomenon, caused by photoelectric effect, electron together with ion attachment and secondary electron emission, will react on surrounding plasma, changing the density distribution and temperature of electrons and ions. Therefore, when an asteroid has rotational velocity, day-night alternation always exists, the equilibrium of interaction between asteroids and surrounding plasma will be constantly reconstructed, efficient dynamic analysis of surface charging is necessary.

Surface charging of asteroids belongs to the field of interaction between solar wind and asteroids. Researches in this direction can be roughly divided into three categories: electrification properties and electrostatic driving dynamics of surface dust particles, overall charging properties of asteroid surfaces, and deep charging of asteroids induced by surface charging. Existing researches center on the charged properties \citep{2016JGRE..121.2150Z,2019ITPS...47.3710O} and floating motion \citep{hartzell2013dynamics} of regolith grains on the surface, mostly using magnetohydrodynamic (MHD) or hybrid particle-in-cell (Hybrid PIC) model. Here we focus on the second category. At the terminator between night and day, the maximum electric field strength appears, almost 10 times that of dayside \citep{2023ApJ...952...61X}, in this region, electron temperature is the most important parameter \citep{2014P&SS...90...10S}. On the dayside of asteroid, photoelectron current usually dominates, surface potential is always positive, theoretical possibilities have been discussed \citep{2020Ap&SS.365...23K}. Correspondingly, on the night side, charging current mainly comes from the incidence of electrons and ions in solar wind plasma, the surface presents a negative potential, during solar energetic particle (SEP) events, this negative potential may reach thousands \citep{2007GeoRL..34.2111H}. However, if enough secondary electrons are emitted, potential of night side can sometimes become positive \citep{halekas2009lunar}.

While the theoretical foundation of surface charging is relatively solid, static simulations of asteroids at specific angles are also feasible, a new grid-free 2D plasma simulation code has been invented \citep{2014Icar..238...77Z}, make it able to discuss interaction between solar wind and a small, irregularly shaped asteroid. But how to dynamically model and verify the surface potential of asteroids with rotation remains an urgent problem to be solved. Operating traditional modeling techniques for three-dimensional simulation means that after asteroid rotates a certain angle, surrounding environment and surface potential of the entire asteroid need to be recalculated through integration and solving ordinary differential equations, resulting in a significant rise in computational complexity, analysis of potential under different rotation periods will be extremely challenging.

In this paper, neural network is used to calculate surface potential of asteroid under given solar wind and material parameters rather than numerical calculation, which is adopt to form the training dataset, greatly improving the efficiency of real-time analysis. We use COMSOL Multiphysics to perform multi-scale modeling simulation, namely divide the whole simulation area with a radius of 500 m into several scales, adopt the smallest scale, i.e. the highest accuracy for the region around asteroid, then increase simulation scale for external plasma, thereby reducing computational complexity. We set three major conditions in order to comprehensively analyze how rotation of asteroids affect their charging properties. Specifically, we first explore asteroids exposed to typical solar wind with fixed surface material, rotation periods from 1 hour to half a year are all considered. Then solar storm is taken into consideration. A coronal mass ejection (CME) existing in early May, 1998 product rapid changes in solar wind density and temperature, four stages of this event have their distinct characteristics, further make vastly separated potential. Due to the complex surface composition of asteroids, it is necessary to explore the charging properties of different minerals, plagioclase, orthopyroxene and ilmenite were studied.

Specific surface charging mechanism and composition of the current balance equation are introduced in Section \ref{sec:MSC}, including photoelectron current, incident electron and ion current, secondary electron current, backscattered current, and conduction current in a thin range of asteroid's surface. How surface potential react on surrounding plasma will also be described in this Section. Establishment of model and parameter settings are shown in Section \ref{sec:MD}. Comparison and analysis of simulation results are present in Section \ref{sec:SR}. Finally, we summarize our work in Section \ref{sec:Con}.

\section{Mechanism of Surface Charging} \label{sec:MSC}
\subsection{Current balance equation}
The current balance equation is the basis of charging theory, surface potential attribute to the accumulation of surface charges, caused by interaction between asteroid and space environment. It is feasible to construct the equilibrium equation in the case of treating a thin area on the surface of asteroid as a dielectric \citep{Synchronization}.

The current balance equation can be expressed as following:

\begin{equation}
C \frac{d U}{d t}=-J_{e}+J_{i}+J_{s e}+J_{s i}+J_{p h}+J_{b s e}-J_{c}
\label{eq:e1}
\end{equation}
where $\mathit{U}$ is instantaneous surface potential of asteroid, $\mathit{C}$ is equivalent capacitance, determined by relative permittivity $\mathit{\varepsilon_r}$. $\mathit{J_e}$ and $\mathit{J_i}$ are environmentally induced electron and ion current density, respectively, $\mathit{J_{s e}}$ and $\mathit{J_{s i}}$ are secondary emitted electron current density caused by electron and ion,  $\mathit{J_{p h}}$ is photoelectron current density, $\mathit{J_{b s e}}$ is backscattered electron current density, and $\mathit{J_c}$ is conduction current density.

In order to solve Equation (\ref{eq:e1}), necessary expressions of each current density are presented below.

1. Photoelectron current emitted from asteroid is a function composed of locational material, solar radiation flux $\mathit{S(E)}$, and electron yield of a single photon $\mathit{W(E)}$ \citep{2004sei..book.....H}. For surfaces with zero or negative potential, photoelectrons will successfully escape, creating initial photoelectron current density:

\begin{equation}
J_{p h 0}=\int_{W_{0}}^{\infty} W(E) S(E) d E
\label{eq:e2}
\end{equation}
where $\mathit{W_0}$ is work function.

Considering that positive surface potential hinders the overflow of photoelectrons, the expression for photoelectron current can be revised to:

\begin{equation}
J_{p h}=\left\{\begin{array}{l}
J_{p h 0}, U \leq 0 \\
J_{p h 0} \exp \left(\frac{-e U}{k T_{p h}}\right)\left(1+\frac{e U}{k T_{p h}}\right), U>0
\end{array}\right.
\label{eq:e3}
\end{equation}
where $\mathit{e}$ is the unit charge, $\mathit{k}$ is Boltzmann constant, $\mathit{T_{p h}}$ is photoelectron temperature, which equals 2.2eV in this paper.

2. Assuming that both electrons and ions follow bi-Maxwellian distribution \citep{2008AdSpR..42.1307N}, when surface potential $\mathit{U\leq 0}$, only the fraction of electron with $\mathit{E\geqslant eU}$ can arrive the surface \citep{7407629}, that means the original energy of electron corresponds to $\mathit{E-eU}$. Therefore, incident electron current density is \citep{1981RPPh...44.1197W}:

\begin{equation}
J_{e}=\left\{\begin{array}{l}
\sum_{j=1}^{2} e n_{e, j}\left(\frac{k T_{e, j}}{2 \pi m_{e}}\right)^{\frac{1}{2}} \exp \left(\frac{-e U}{k T_{e, j}}\right), U \leq 0 \\
\sum_{j=1}^{2} e n_{e, j}\left(\frac{k T_{e, j}}{2 \pi m_{e}}\right)^{\frac{1}{2}}\left(1+\frac{e U}{k T_{e, j}}\right), U>0
\end{array}\right.
\label{eq:e4}
\end{equation}
where $\mathit{n_e}$ is electron density, $\mathit{T_e}$ is electron temperature, $\mathit{m_e}$ is Electron Mass.

In the similar manner, incident ion current density can be described as:

\begin{equation}
J_{i}=\left\{\begin{array}{l}
\sum_{j=1}^{2} q_{i} n_{i, j}\left(\frac{k T_{i, j}}{2 \pi m_{i}}\right)^{\frac{1}{2}}\left(1+\frac{q_{i} U}{k T_{i, j}}\right), U \leq 0 \\
\sum_{j=1}^{2} q_{i} n_{i, j}\left(\frac{k T_{i, j}}{2 \pi m_{i}}\right)^{\frac{1}{2}} \exp \left(\frac{-q_{i} U}{k T_{i, j}}\right), U>0
\end{array}\right.
\label{eq:e5}
\end{equation}
where $\mathit{q_i}$ is ionic charge, $\mathit{n_i}$ is ion density, $\mathit{T_i}$ is ion temperature, $\mathit{m_i}$ is ion mass.

3. Secondary electron current density due to electrons is represented as the product of secondary electron yield and incident electron current density:

\begin{equation}
J_{s e}=\left\{\begin{array}{l}
Yse(E) J_{e}, U \leq 0 \\
Yse(E+e U) J_{e}, U>0
\end{array}\right.
\label{eq:e6}
\end{equation}

That due to ions is in the same form:

\begin{equation}
J_{s i}=\left\{\begin{array}{l}
Ysi(E-e U) J_{i}, U \leq 0 \\
Ysi(E) J_{i}, U>0
\end{array}\right.
\label{eq:e7}
\end{equation}

Backscatter current is the current generated by surface scattering of incident electrons, which is determined by backscattered electron yield of the surface material:

\begin{equation}
J_{b e}=\left\{\begin{array}{l}
Y b e(E) J_{e}, U \leq 0 \\
Y b e(E+e U) J_{e}, U>0
\end{array}\right.
\label{eq:e8}
\end{equation}

where $\mathit{Yse}$ and $\mathit{Ysi}$ are secondary electron yield induced by electrons and ions respectively, $\mathit{Ybe}$ is backscattered electron yield, their formulas are adopted according to reference \citep{Impact}.

4. Conduction current represents the current conducted from the surface of equivalent dielectric to the back:

\begin{equation}
J_{c}=\frac{U}{\rho d}
\label{eq:e9}
\end{equation}
where $\mathit{\rho}$ is conductivity, $\mathit{d}$ is the thickness of dielectric, here it equals 0.01mm.

\subsection{Interaction with Solar Wind Plasma}
\label{subsec:ISW}
When surface charging appears, electric field is formed around the asteroid, electrons and ions will move under the traction of electric field force, drift-diffusion equations \citep{2020PhPl...27a3505G} are widely used to solve these problems of plasma physics.

\begin{equation}
\left\{\begin{array}{l}
\frac{\partial n_{e}}{\partial t}+\nabla \cdot \boldsymbol{\mathbf\Gamma}_{e}=R_{e}-(\mathbf{u} \cdot \nabla) n_{e} \\
\boldsymbol{\mathbf\Gamma}_{e}=-\mu_{e} \mathbf{E} n_{e}-D_{e} \nabla n_{e}
\end{array}\right.
\label{eq:e10}
\end{equation}
where $\mathit{R_e}$ is electron generation rate, $\mathit{\mathbf\Gamma_e}$ is the related flux-density vector, $\mathit{\mathbf{u}}$ is the velocity vector, $\mathit{\mu_{e}}$ and $\mathit{D_e}$ are the mobility and the diffusion coefficient of electrons, satisfy Einstein relation, can be represented as:

\begin{equation}
\left\{\begin{array}{l}
\mu_{e}=\frac{e}{m_{e} v_{c e}} \\
D_{e}=\frac{k T_{e}}{m_{e} v_{c e}}
\end{array}\right.
\label{eq:e11}
\end{equation}
where $\mathit{v_{ce}}$ is the frequency of electron-neutral collisions, can be derived from kinetic theory of plasma \citep{1978npi..book.....G,1970pewp.book.....G}.

Representing the frequency of ion-neutral collisions with $\mathit{v_{ci}}$, the mobility and the diffusion coefficient of ions can be expressed analogously:

\begin{equation}
\left\{\begin{array}{l}
\mu_{i}=\frac{q_{i}}{m_{i} v_{c i}} \\
D_{i}=\frac{k T_{i}}{m_{i} v_{c i}}
\end{array}\right.
\label{eq:e12}
\end{equation}

Changes in surrounding plasma, especially density and temperature of electrons and ions, will alter surface potential of asteroid, then modify electric field distribution in this area, causing the plasma to undergo changes again, forming a cycle. During this process, the rotation of asteroid and day-night alternation on the surface will greatly affect the equilibrium value. Therefore, to accurately analyze surface charging phenomenon, previous static analysis is insufficient, we need to dynamically simulate the asteroid and surrounding plasma.

\section{Model Description} \label{sec:MD}
\subsection{Neural Network}
Substitute solar wind and material parameters into the current balance equation, it can transform into an ordinary differential equation solely related to potential. Potentials obtained from solving this equation, along with the parameters, will build a dataset to train a BP (Back Propagation) neural network \citep{2023ITPS...51.1181Q}, for the reason that achieve fast and accurate calculation of surface potential in subsequent dynamic simulation.

\begin{figure}[ht!]
\centering
\includegraphics[width=0.4\linewidth]{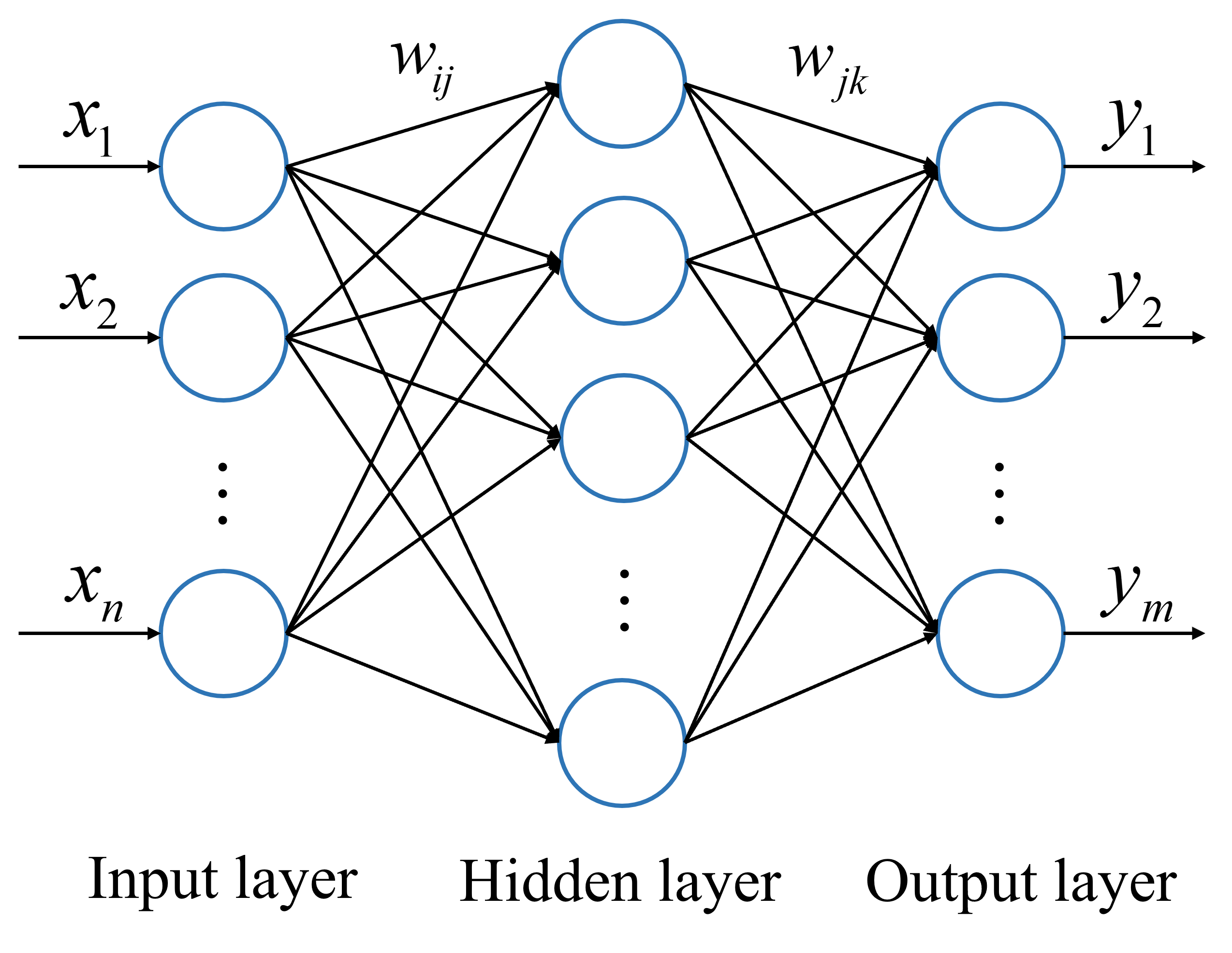}
\caption{Structure of BP neural network. $x_i$ represents input data while $y_i$ represents output data. $w_{ij}$ and $w_{jk}$ mean neurons' weights, the goal of training is to update them then minimize the loss function.
\label{fig:Fig1}}
\end{figure}

BP neural network is a multi-layer feedforward network trained by error inverse propagation algorithm \citep{zhu2023method}. Its topology, as shown in Figure \ref{fig:Fig1}, can be divided into three levels: input layer, hidden layer, and output layer \citep{adil2022effect}. Take mean square error (MSE) as standard, by learning the input-output dataset with steepest gradient descent method, through reverse propagation to continuously adjust weights of each layer, in an effort to minimize errors, it can structure and store massive mapping relationships hidden under a big dataset, owning excellent generalization ability and strong fault tolerance \citep{2022MRE.....9b5504L}.

Backpropagation algorithm has two elements, in simple terms, forward propagation of input information and reverse propagation of output errors. For the former element, input parameters pass through hidden layer with weights and biases, then undergo nonlinear transformation, generating output potential. This output will be compared with the true value in dataset, then obtain loss function, the purpose of training is to minimize it. If the fitness of network cannot meet requirements, commonly indicates large loss function, we need to adjust the mapping relationship in neural network. During reverse propagation, errors will be transmitted back through hidden layer to input layer, and distributed to all neurons in each layer, then network can modify weights and biases with Gradient Descent Algorithm \citep{eker2023comparison} according to errors of each neuron \citep{qian2023accuracy}. This propagation will be repeated until network achieve desired accuracy.

\begin{table}[ht!]
\caption{Parameters of neural network}
\tabletypesize{\scriptsize}
\tablewidth{0pt}
    \centering
    \begin{tabular}{c c c}
    \toprule
         \multirow{12}*{Input}& \multirow{5}*{Solar wind plasma}&Electron density  \\
         ~&~&Ion density  \\
         ~&~&Electron temperature  \\
         ~&~&Ion temperature  \\
         ~&~&Solar radiation  \\
         \cline{2-3}
         ~&\multirow{5}*{Material}&Work function  \\
         ~&~&Maximum secondary electron yield  \\
         ~&~&Energy for maximum secondary electron yield  \\
         ~&~&Relative permittivity  \\
         ~&~&Conductivity  \\
         \cline{2-3}
         ~& \multicolumn{2}{c}{Current potential (instantaneous only)}  \\
         ~& \multicolumn{2}{c}{Time step (instantaneous only)}  \\
         \midrule
         \multirow{2}*{Output}& \multicolumn{2}{c}{Potential at the next moment (instantaneous)} \\
         ~& \multicolumn{2}{c}{Equilibrium potential(equilibrium)} \\
         \bottomrule
    \end{tabular}
    \label{tab:Tab1}
\end{table}

All input and output parameters selected are present in Table \ref{tab:Tab1}. In this paper, two neural networks are trained, in order to calculate instantaneous potential and equilibrium potential, respectively. The former is applied to situations where the time step is less than 1s, mainly for the beginning of simulation. When the period is long enough to make time step reaches more than 1s, much higher than the time needed for surface to reach equilibrium potential, we can use equilibrium potential for calculation.

For the aim of improving the universality and accuracy of neural networks, dataset includes random combinations of various material parameters and solar wind plasma parameters from standard condition to extreme, such as solar wind number density changes between $\mathit{\rm 5\times10^6/m^3}$ and $\mathit{\rm 5\times10^9/m^3}$, electron temperature ranging from 5eV to 20eV. Thus, neural networks can so reliably establish the mapping relationship between input parameters and output potential, making them suitable for a wide range of applications. Since the network used to calculate instantaneous potential depends heavily on the previous potential and time step, to fully consider the impact of these two parameters, we adopt 879950 pieces of data for training, each piece of data contains 12 input parameters and 1 output parameter. The specific parameters are listed in Table \ref{tab:Tab1}, far more than the that calculating equilibrium potential, which uses 17970 pieces of data, each piece of data contains 1- input parameters and 1 output parameter.

Before training, data normalization is considered, eliminate differences in order of magnitude between parameters, thus avoid unnecessary network errors. ReLU is used as activation function, it can compensate for the problem of Vanishing Gradient that Sigmoid and Tanh have \citep{kiliccarslan2023novel}, helping network learn complex patterns in data.

\subsection{Dynamical Multi-scale Model}
A three-dimensional model has been developed to investigate surface charging phenomenon in solar wind plasma, we use COMSOL Multiphysics to realize the simulation. COMSOL is a universal software platform used for modeling and simulating physical field problems. It is based on finite element analysis, capable of exploring real physical phenomena by simulating a single physical field and flexibly coupling multiple physical fields.

\begin{figure}[ht!]
\centering
\includegraphics[width=0.6\linewidth]{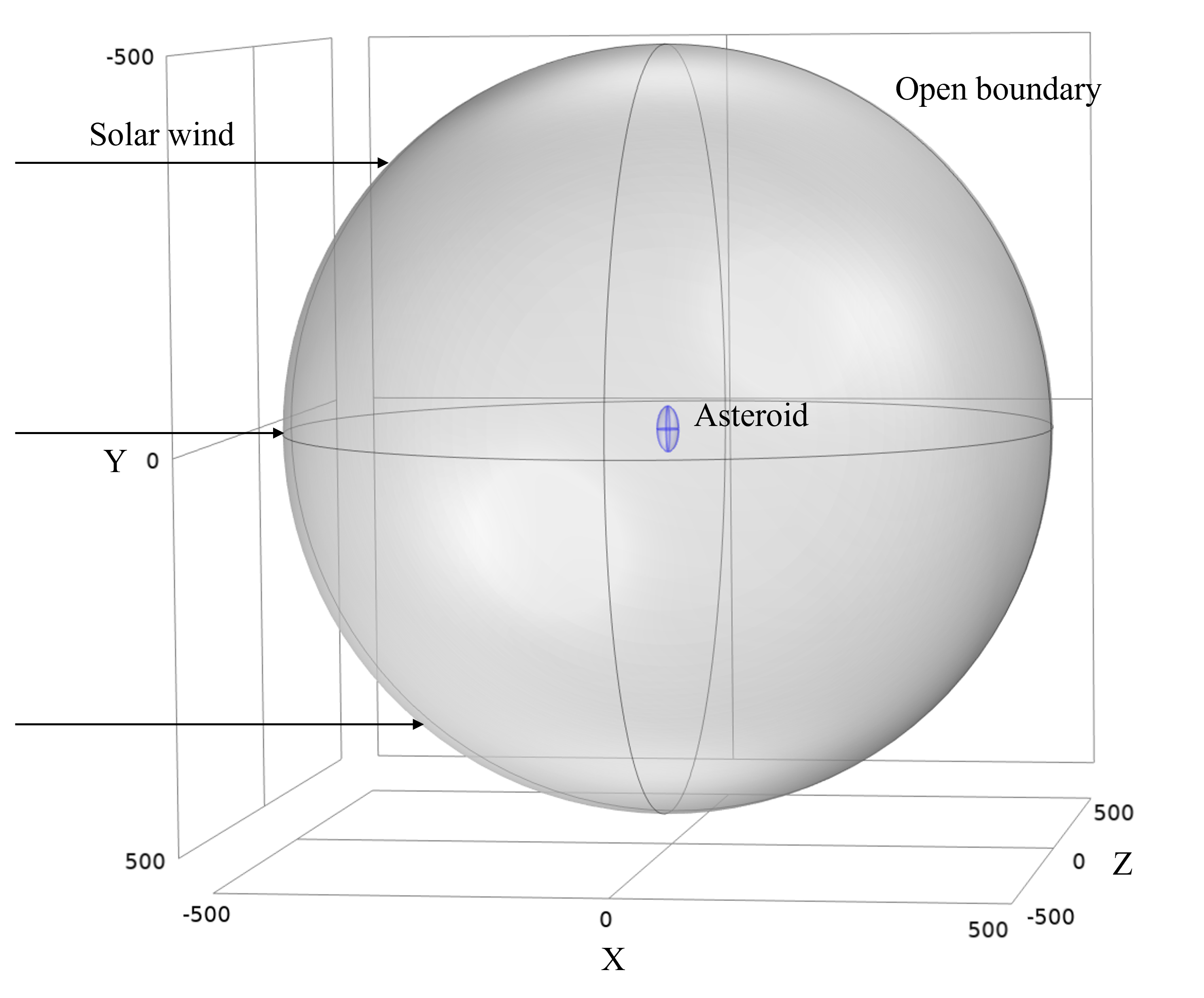}
\caption{The schematic diagram of the simulation geometries. The gray sphere indicates our simulation domain and the blue ellipsoid represents asteroid. The three black arrows on the left show the direction of solar wind.
\label{fig:Fig2}}
\end{figure}

We use the shape of 2016HO3 with a width-to-length ratio smaller than 0.48 to represent asteroid, regard it as a triaxial ellipsoid with semi-major axis of 29.5, 14, 14m \citep{LI2021114249}. As indicated in Figure \ref{fig:Fig2}, the simulation domain is a sphere with a radius of 500m, asteroid locates at the center of it, the Z-axis is antiparallel to the spin axis of asteroid in Section \ref{subsec:SCP}-\ref{subsec:DAV}, therefore we can simulate scenarios from asteroid's main axis being perpendicular to the solar wind to being parallel to it, while selecting either the X-axis or the Y-axis only allows for considering one of the perpendicular or parallel situations. This spin axis obliquity allows asteroid to pass through some meaningful attitudes during rotation, resulting in the greatest impact of revolution on asteroid's surface potential, it can facilitate comparison and analysis of these special attitudes. However, this obliquity is not common in real situations. To make our simulation more suitable for realistic situations, we studied the effect of obliquity on surface potential by altering the spin obliquity in Section \ref{subsec:EOM}.

Our studies can be summarized as the following steps:

a. Input plenty of solar wind parameters and material parameters, compute instantaneous and equilibrium potentials on asteroid's surface under these conditions with numerical integration, thereby build a dataset.

b. Divide the dataset into training and testing sets, train and optimize neural network.

c. Model asteroid within a simulation domain in Figure \ref{fig:Fig2}, initialize electrostatic field of asteroid's surface and surrounding plasma.

d. Input parameters (including solar wind plasma, solar irradiation, and material) at various locations into neural network, then output surface potential. 

e. Calculate potential and electric field distribution within the remaining domain.

f. Simulate drift-diffusion motion of electrons and ions, then compute surrounding density and temperature of them.

g. Update input parameters of neural network, output surface potential at the next simulation time step.

h. Iterate steps d-g until reaches required duration (e.g., one rotation period of asteroid).

\begin{figure}[ht!]
\centering
\includegraphics[width=0.6\linewidth]{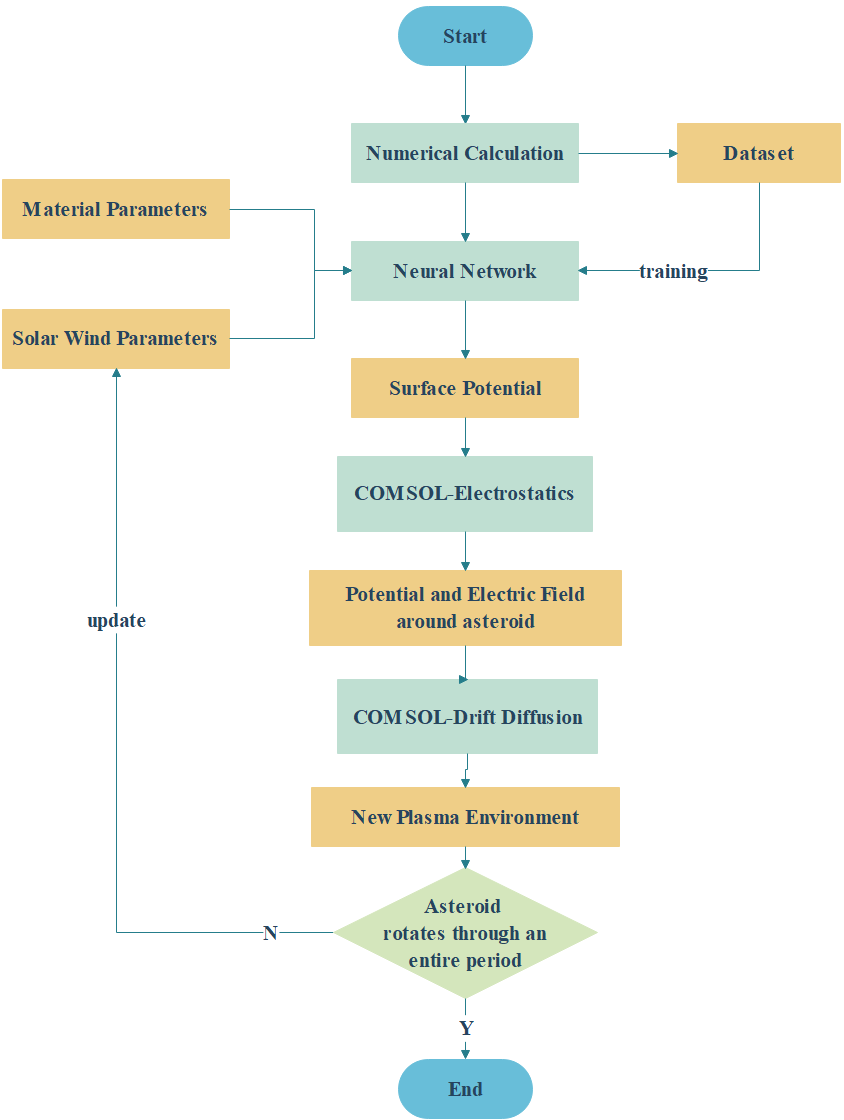}
\caption{Flow chart of modeling and simulation. Numerical calculation only serves to build the dataset, it will not appear in subsequent simulations.
\label{fig:Fig3}}
\end{figure}

This process is illustrated in Figure \ref{fig:Fig3}. It is worth noting that the 1D equations in Section \ref{sec:MSC} are only used for dataset construction and theoretical explanation, which help in building the neural network, accurately complete the mapping from plasma parameters and material parameters to surface potential. In our simulation, we divided the surface of the asteroid into many very small areas. Each area equals a capacitor that has the same surface potential. We use the neural network to calculate the surface potential of the capacitor instead of solving the equations in Section \ref{sec:MSC}. Then we combine all the surface potential of small areas to form the 3D surface potential of the asteroid. After obtaining asteroid's surface potential, the calculations of potential of the remaining simulation domain, electric field around asteroid, and the drift-diffusion motion of charged particles are all done in 3D.

\begin{figure}[ht!]
\centering
\includegraphics[width=0.5\linewidth]{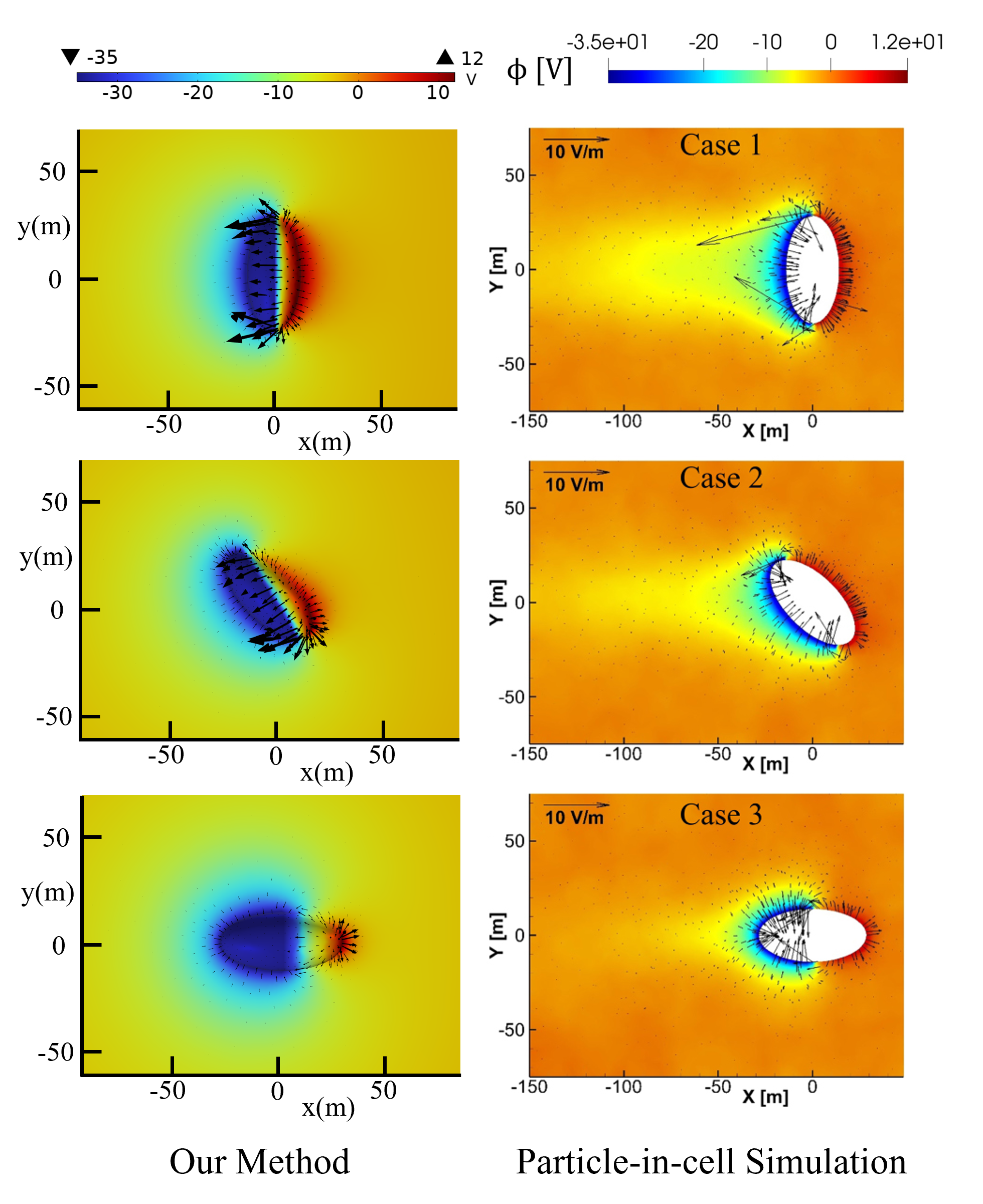}
\caption{Comparison between our simulation method and PIC simulation. The three subfigures on the left shows results of multi-scale modeling simulation, indicating potential and electric field around asteroid. Those on the right are results from reference \citep{2023ApJ...952...61X} under the same circumstances.
\label{fig:Fig4}}
\end{figure}

In order to verify the reliability of our three-dimensional surface charging and particle motion model, we operate parameters of PIC simulation \citep{2023ApJ...952...61X}, where $\mathit{T_e=10}$eV, secondary electron yield has a peak yield of $\mathit{\delta_{\text {max }}=1.0}$ at the energy of $\mathit{E_{\text {max }}=350}$eV. Solar wind parameters are at their typical values, with number density of $\mathit{\rm 5 \times10^6/m^3}$, a velocity of 400km/s, ions and electrons are emitted in the -X direction, time step is 1.0e-6s, while the simulation period is 0.001s. As they assume that the asteroid is static, the angle $\mathit{\alpha}$ between the main axis of asteroid and plane Y-O-Z is fixed. Results are shown in Figure \ref{fig:Fig4}. Notably, under normal condition, to ensure a fair comparison with PIC, we set the plasma near asteroid (about 200m) to be quasi-neutral here, similar to the PIC method \citep{2023ApJ...952...61X}. However, this does not accurately reflect the real situation, due to the solar wind flow and electric fields in space, the potential generally does not drop to zero at 200m from asteroid \citep{2014Icar..238...77Z}. Especially during a solar storm, the Debye length of solar wind plasma can become almost double its normal value. As a result, in simulation of Section \ref{subsec:IRA}, we will consider the plasma to be quasi-neutral 500m away from asteroid, potential around asteroid (about 50m) may still exhibit an obvious negative value.

We can see a positive potential up to +13.3V on the dayside and a negative potential as low as -37.5V on the nightside when the angle $\mathit{\alpha=0^{\circ}}$, while surface potential is about +12V near the subsolar point, and -35V on the nightside with PIC simulation.

\begin{figure}[ht!]
\centering
\includegraphics[width=0.9\linewidth]{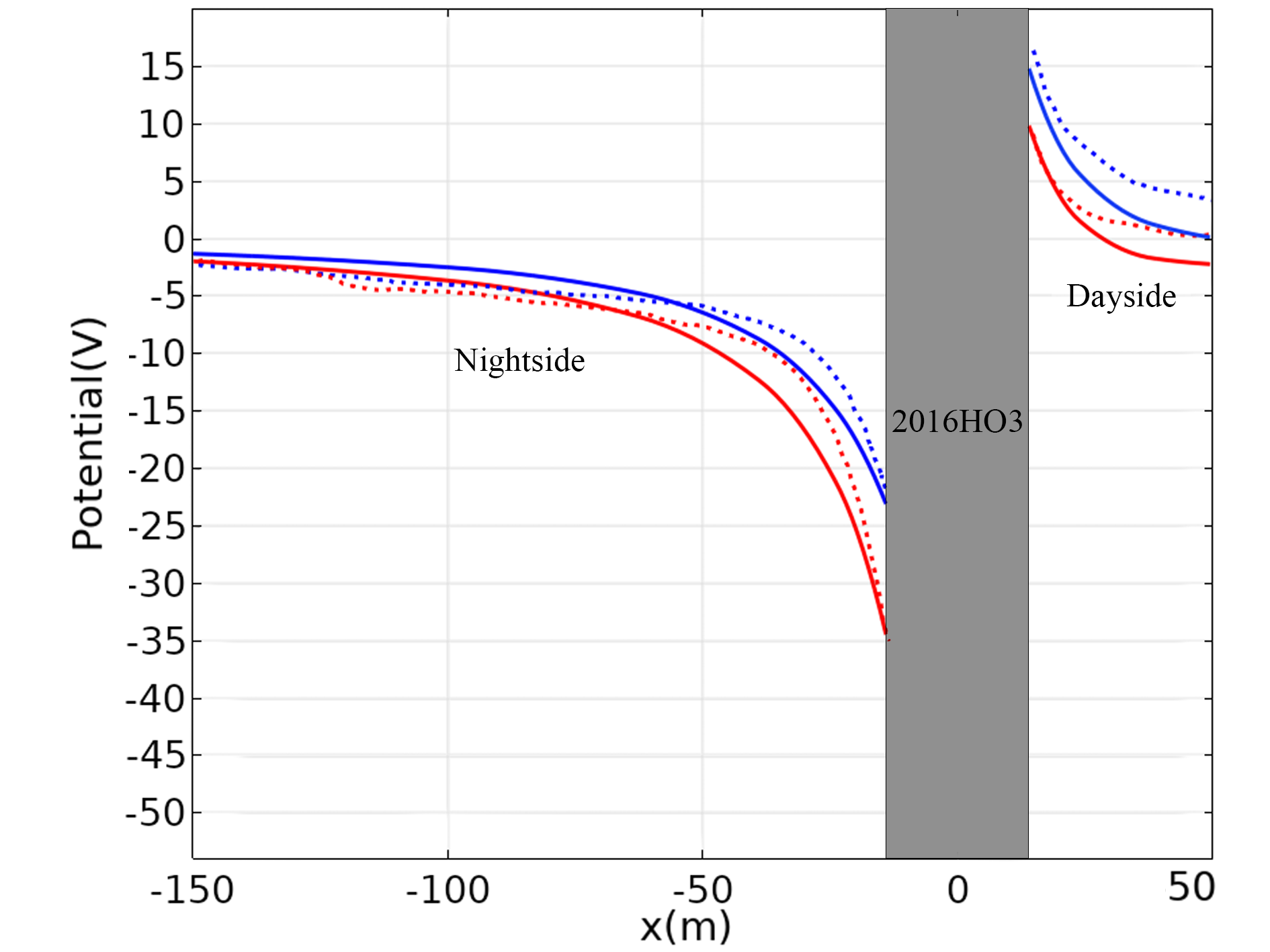}
\caption{Comparison of potential profiles along the X-axis for two cases between our simulation method and PIC simulation. The data are extracted along the X-axis at Y=Z=0, in which the red and blue lines show the results of Cases 4, 8 in reference \citep{2023ApJ...952...61X} respectively.
\label{fig:Fig5}}
\end{figure}

Apart from the three situations mentioned above, we also compared the potential distribution around asteroid for Case 4 and Case 8 in reference \citep{2023ApJ...952...61X}. The difference between simulated parameters in Case 4 and that used for Figure \ref{fig:Fig4} locates at solar wind number density, equals to $\mathit{\rm 1\times10^7/m^3}$ in Case 4, while Case 8 differs from Figure \ref{fig:Fig4}'s simulated environment by $\mathit{\delta_{\text {max }}=3.0}$. Simulated results of our method and PIC simulation with Cases4 and 8 are shown in Figure \ref{fig:Fig5}. Results simulated by our three-dimensional multi-scale model are on the left, and the right subfigure shows results of PIC simulation.

Our results are very close to those of previous research in terms of both curve trend and values. Given that our much larger simulation region than PIC's, we assume that potential 500m away from asteroid is 0. Therefore, within the range -150m$\sim$50m, there exists some minor differences between our method and PIC, which will not exceed 3V. It can be seen that our method is reliable.

\section{Simulation} \label{sec:SR}
We will conduct dynamic simulations asteroids about surface charging under different conditions, the four main categories are shown as follows:

\subsection{Surface Charging Properties of Rotating Asteroids under Normal Condition}
\label{subsec:SCP}

Solar wind flows in the +X direction, number density is equal to $\mathit{\rm 8\times10^6/m^3}$, electron temperature is 15eV while ion temperature is 10eV. Electrons and ions are emitted from the outer boundaries with a bulk speed of 400 km/s, an open boundary condition is applied here, where all particles can escape through the boundary. Small-scale mesh is used within 50 m around asteroid for fine simulation, as to the remaining areas, mesh adopt will scale up appropriately in line with standard of research.

Solar radiation received by surface depends on solar elevation angle, solar radiation reaches the peak at the subsolar point, then reduce gradually along with decreasing solar elevation angle till cross the terminator, solar radiation intensity thus become 0 on the night side. Plagioclase regards as material on the surface of asteroid, its work function is 5.58eV, threshold wavelength is 238nm. Secondary electron emission is taken into account here, with maximum secondary electron yield of 2.8 at the energy of 1000eV. The simulation period is set in accordance with rotation period, initial time step is 1e-5s, then become 1s for asteroids have short rotation periods, such as 1 hour, finally grow to 1 hour when rotation period is half a year. However, no matter how long rotation period is, time step always much smaller than it, making it reasonable to analyze potential under various rotation angles.

Surface potential of asteroids with eight rotation periods from 1 hour to half a year are simulated, even without solar storm, there can be macroscopic potential differences on surface of asteroids with different periods.

\begin{figure}[ht!]
\centering
\includegraphics[width=0.7\linewidth]{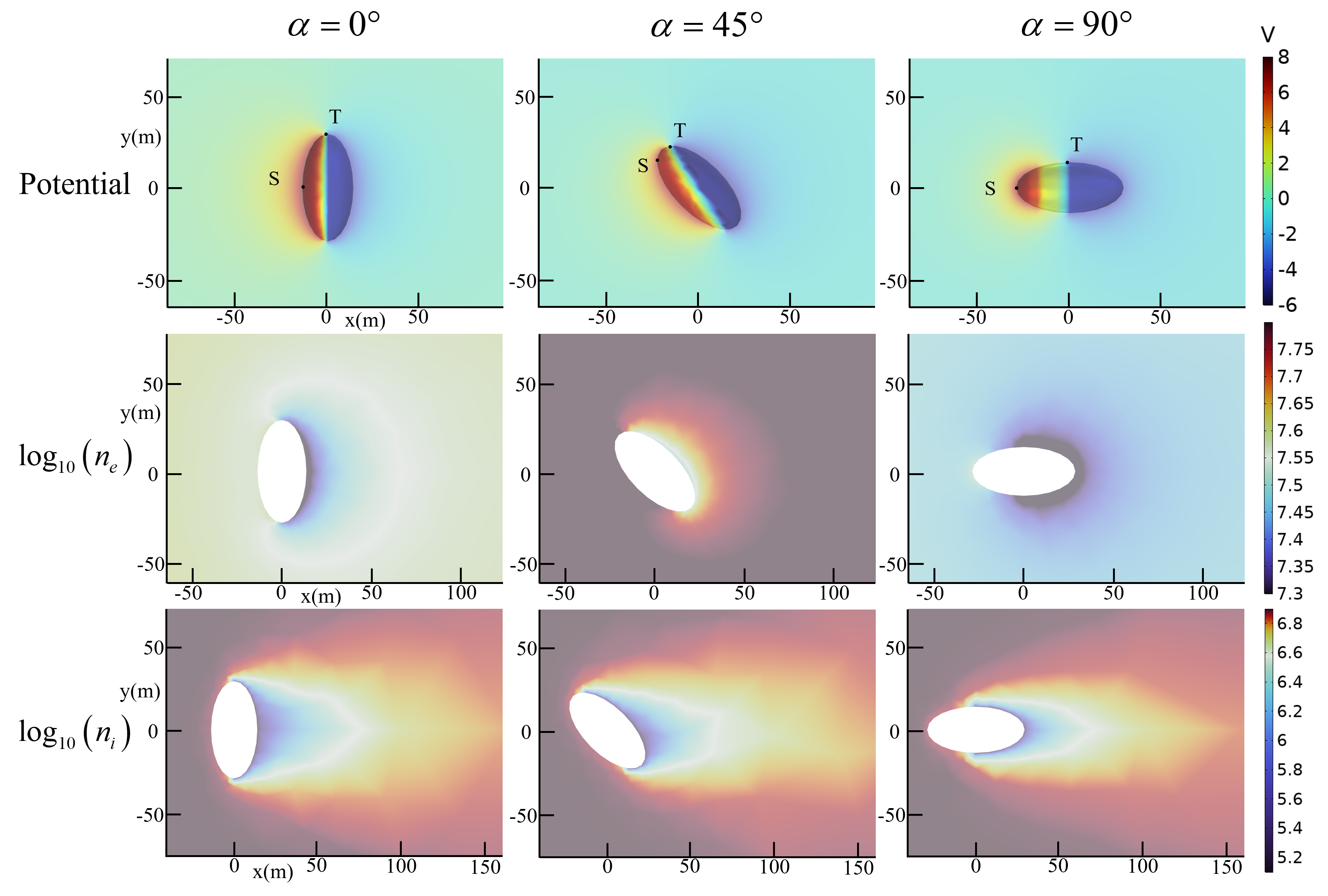}
\caption{Surface potential and surrounding plasma of asteroid with rotation period of 1 day. Subfigures in the first row show surface potential of asteroid, points $S$ and $T$ locate at the subsolar point and the terminator. The other two rows represent densities of electrons and ions around asteroid in the X-O-Y plane respectively.
\label{fig:Fig6}}
\end{figure}

Figure \ref{fig:Fig6} displays the surface potential of asteroid with three classical attitudes, as well as the distribution of electron density and ion density around it, the rotation period is always 1 day. Consistent with the results of static simulation, the minimum potential the asteroid can achieve on the nightside, -5.74V, is the lowest when $\mathit{\alpha=45^{\circ}}$ compared with other two conditions, -4.6V when $\mathit{\alpha=0^{\circ}}$ and -4.5V when $\mathit{\alpha=90^{\circ}}$. Meanwhile, if the main axis of asteroid is perpendicular to the direction of solar wind incidence, potential on the dayside is obviously higher than other situations.

As for the surrounding solar wind plasma, because of its much higher mobility and diffusion coefficient, electron responds more promptly to the electric field, has a relatively uniform distribution, there won't be exaggerated faults for density. Correspondingly, for ions primarily driven by the general flow of solar wind, will form a visible plasma wake at the tail of asteroid. As a result, during the rotation of asteroid, electron density undergoes significant changes with the surface charging conditions. Electrons entering the region will accumulate around asteroid, with a peak value up to $\mathit{\rm 8\times10^7/m^3}$, 10 times the initial solar wind number density. At the same time, ion density will only fluctuate within a small range, even cannot exceed $\mathit{\rm 1\times10^7/m^3}$, and exhibits an extremely low value in the wake of asteroid. This phenomenon leads to differences in surface potential of asteroids during rotating.

\begin{figure}[ht!]
\centering
\begin{subfigure}
{\includegraphics[width=0.75\linewidth]{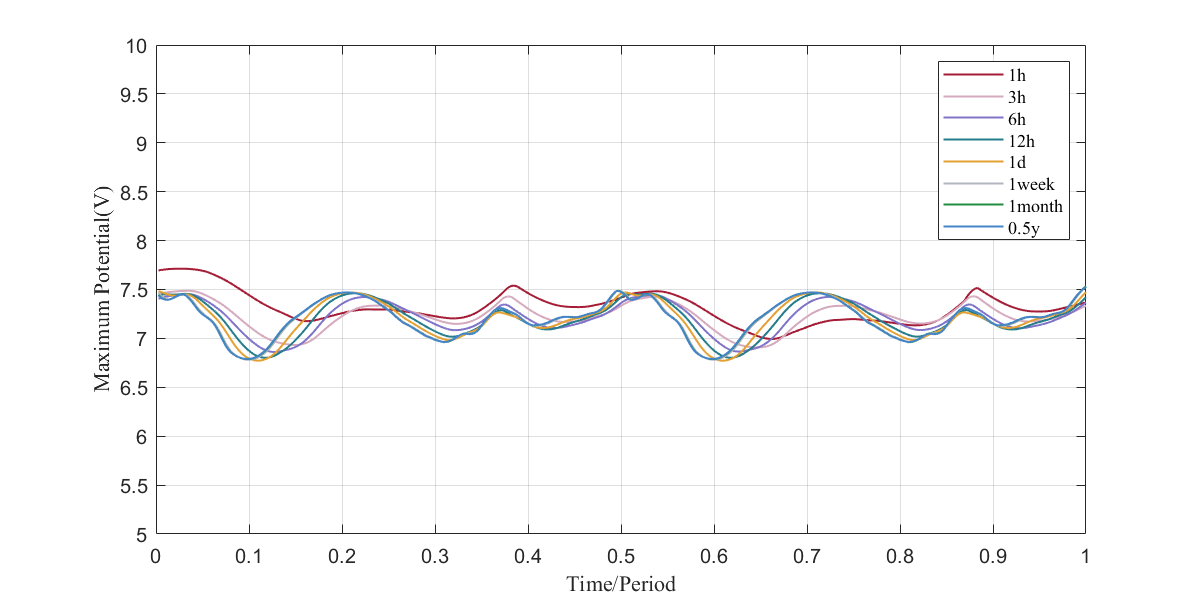}
}
\end{subfigure}

(a)

\begin{subfigure}
{\includegraphics[width=0.75\linewidth]{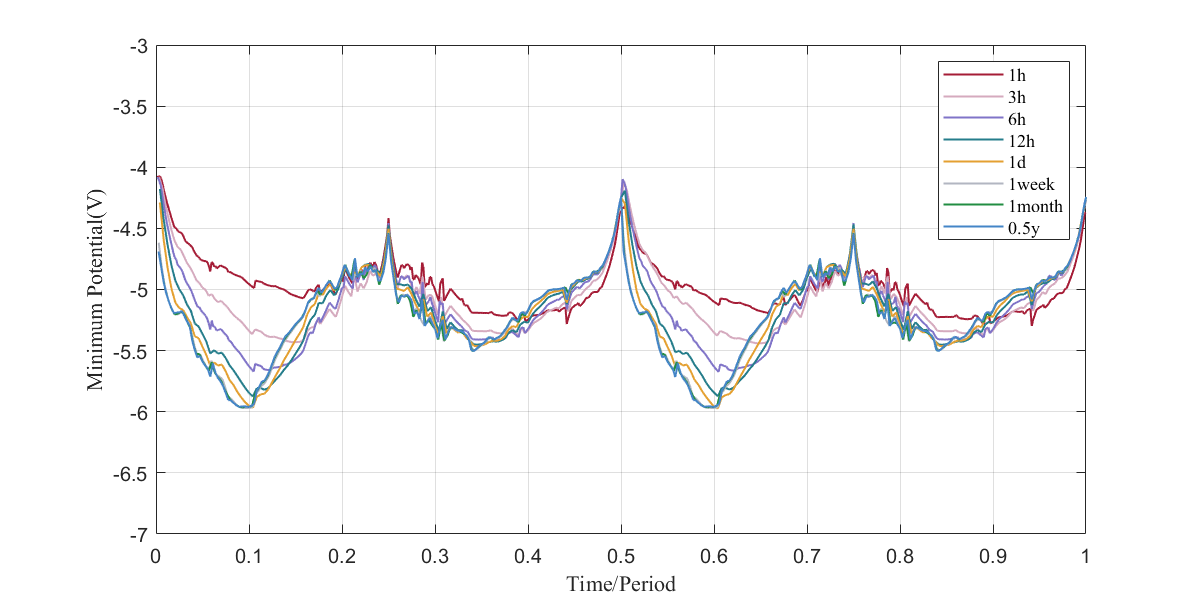}
}
\end{subfigure}

(b)

\caption{Potential under different rotation periods. The initial attitude of asteroid is the same as that in Figure \ref{fig:Fig2}, the major axis of asteroid is normal to the direction of solar wind incidence, i.e. $\mathit{\alpha=0^{\circ}}$. (a) shows the maximum potential on the surface of asteroid within one rotation period, (b) shows the minimum potential.
\label{fig:Fig7}}
\end{figure}

The maximum and minimum potential on the surface of asteroid with different rotation periods are shown in Figure \ref{fig:Fig7}. Due to the fact that photoelectron current is the main source of surface charging on the dayside, potential here is less affected by fluctuations in electron and ion density compared to that on the nightside. The maximum potential always locates at the point of direct sunlight, appears much smoother and more stable than the minimum potential. However, their general downward trend with the growth of rotation period is still consistent.

Asteroids with short rotation periods, such as 1 hour, have fast rotation angular speed, bring rapid alternation of illuminated face and shadowed face, along with electric field. Its ability to collect electrons is rather weak, potential has no clear valley value during rotation. Later, as the period increases, before reaching 1 week, the charge accumulation around asteroid will gradually become obvious. The peak value of average electron density in small-scale area will increase from $\mathit{\rm 3.07\times10^7/m^3}$ at 1 hour to $\mathit{\rm 7.60\times10^7/m^3}$ at 1 week, thus the minimum potential continues to decrease, in the meantime reduce the maximum potential. The range of potential can be changed from +6.99V$\sim$+7.71V and -5.30V$\sim$-4.07V at 1 hour to +6.78V$\sim$+7.53V and -5.97V$\sim$-4.24V at 1 week.

Apart from period, the direction and angle of rotation can also affect charging results, for example, $\mathit{\alpha=45^{\circ}}$ and $\mathit{\alpha=135^{\circ}}$ are the same situation for asteroids when we use static simulation, but considering that they are rotating, their results differ significantly. Besides, from Figure \ref{fig:Fig7}, we can conclude that no matter how long the period is, surface potential tends to be low when solar wind is obliquely incident, this phenomenon will become more obvious as rotation period increases. That's because when asteroid rotates to $\mathit{\alpha=25\sim45^{\circ}}$, the subsolar point $\mathit{S}$, where often has the maximum potential, is close to the terminator $\mathit{T}$. According to Poissons Equation, the sudden change in potential over a short distance will strengthen electric field near asteroid, accelerate electrons to $2.21\times10^7$m/s, the resulting higher electron density and electron temperature therefore reduce surface potential. This process continues until electric field and plasma reach a dynamic balance, which will be broken and reshaped after asteroid rotates through a large enough angle. 

This explains why the curve of a short-period asteroid is rather stable: before surface potential falls low enough, asteroid and plasma must move toward a new balance. In this new situation, electric field is much softer, unable to prevent electrons from escaping due to density difference. In the same principle, asteroids with periods longer than 1 week possess the capacity to attain equilibrium when solar wind is incident obliquely, -5.97V is the minimum potential asteroid can reach with these given parameters, it is also the potential that enables plasma and asteroid to become balanced when $\mathit{\alpha=36^{\circ}}$.

\subsection{Impact of Rotation on Asteroids during the Passage of Solar Storm}
\label{subsec:IRA}

On 29 April 1998, a CME emitting from the Sun was observed by The Lunar Prospector (LP) spacecraft, having a significant impact at Earth \citep{2012JGRE..117.0K04F}. A forward shock occurred at 21 UT on May 1, followed by a clear CME, with counterstreaming electrons and a depressed proton temperature, then multiple shocks and CMEs appeared, finally came to an end at 1 UT on May 4 \citep{1999GeoRL..26..161S}. During the passage of CMEs, both density and temperature of solar wind will undergo drastic changes, even obtain values 10 times that in usual.

This entire series of events can be divided into four distinct temporal parts: typical solar wind; a dense, hot shock; early stage of CME; and late stage of CME \citep{2012JGRE..117.0K03Z}. Four simulation cases with their solar wind parameters have been performed in Table \ref{tab:Tab2}.

\begin{table}[ht!]
\caption{Plasma parameters during passage of solar storm}
\tabletypesize{\scriptsize}
\tablewidth{0pt}
    \centering
    \begin{tabular}{ccccc}
    \toprule
         Stage&Typical&	Shock&	Early CME&	Late CME \\
         \midrule
         Number density($\mathit{\rm /cm^{-3}}$)&	5$\sim$8&	20&	3&	$\geqslant$50 \\
         Bulk speed(km/s)& 400$\sim$450&	600&	650&	500 \\
         $\mathit{T_e}$(eV)& 15&	80&	9&	16 \\
         $\mathit{T_i}$(eV)& 10&	40&	6&	4 \\
         $\mathit{c_s}$(km/s)& 37.91&	87.54&	29.36&	39.15 \\
         \bottomrule
    \end{tabular}
    \label{tab:Tab2}
\end{table}

Surface mineral of asteroid is plagioclase, in terms of researches about charging properties, we set a rotation period of 0.467 hour in each case, equivalent to the spin period of 2016 HO3. In the discussion of how rotation period effect on surface charging, other periods are added.

\begin{figure}[ht!]
\centering
\includegraphics[width=0.9\linewidth]{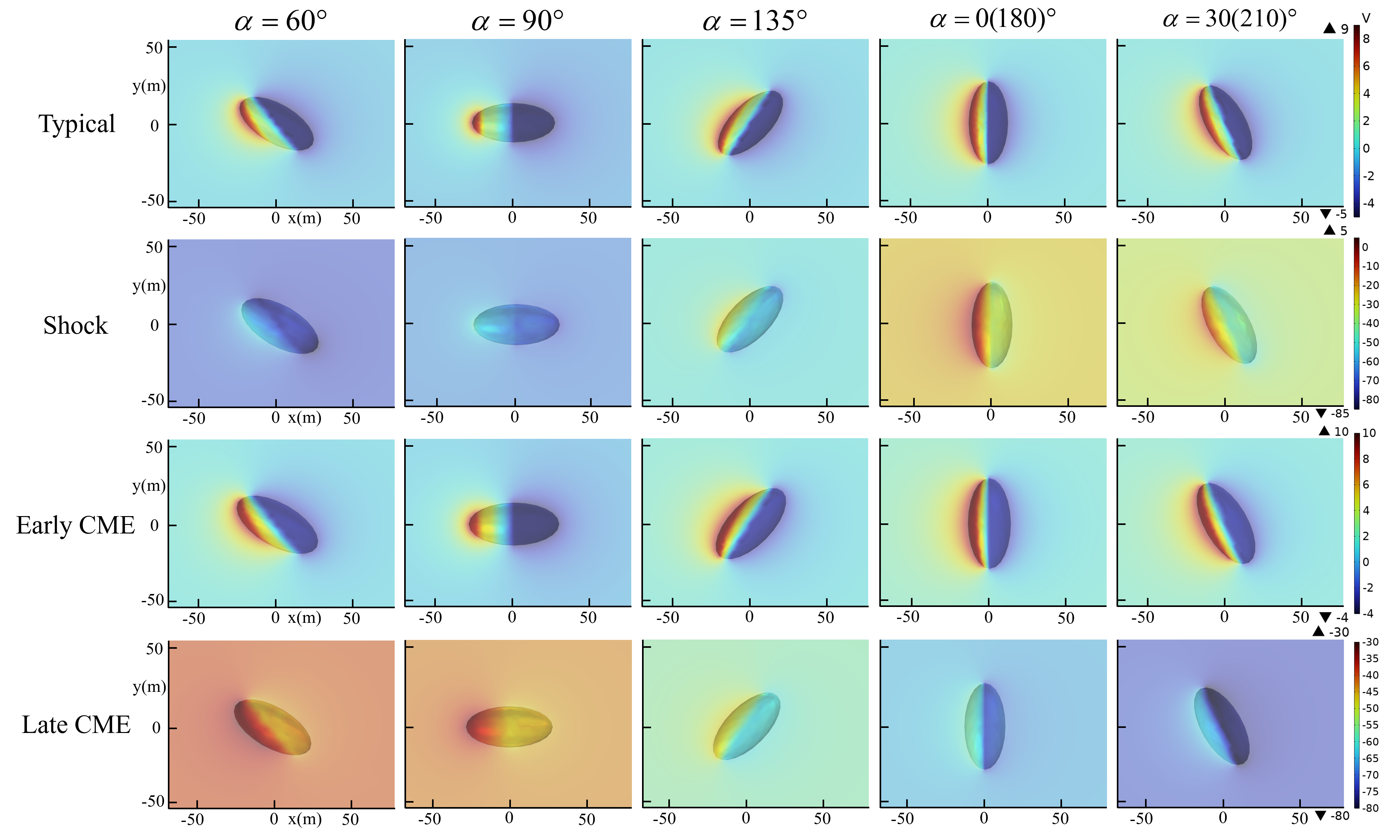}
\caption{Simulated potential during passage of the modeled CME. Potentials of X-O-Y plane and asteroid's surface are displayed. $\alpha$ is the angle between solar wind and the major axis of asteroid, ranging from 0 to 180 degrees. $\alpha=180^{\circ}/210^{\circ}$ is equivalent to $\alpha=0^{\circ}/30^{\circ}$. To clearly present surface potential of asteroid, we only show results 50m around asteroid. Plasma 500m away from asteroid is assumed to be quasi-neutral.
\label{fig:Fig8}}
\end{figure}

\begin{figure}[ht!]
\centering
\includegraphics[width=0.95\linewidth]{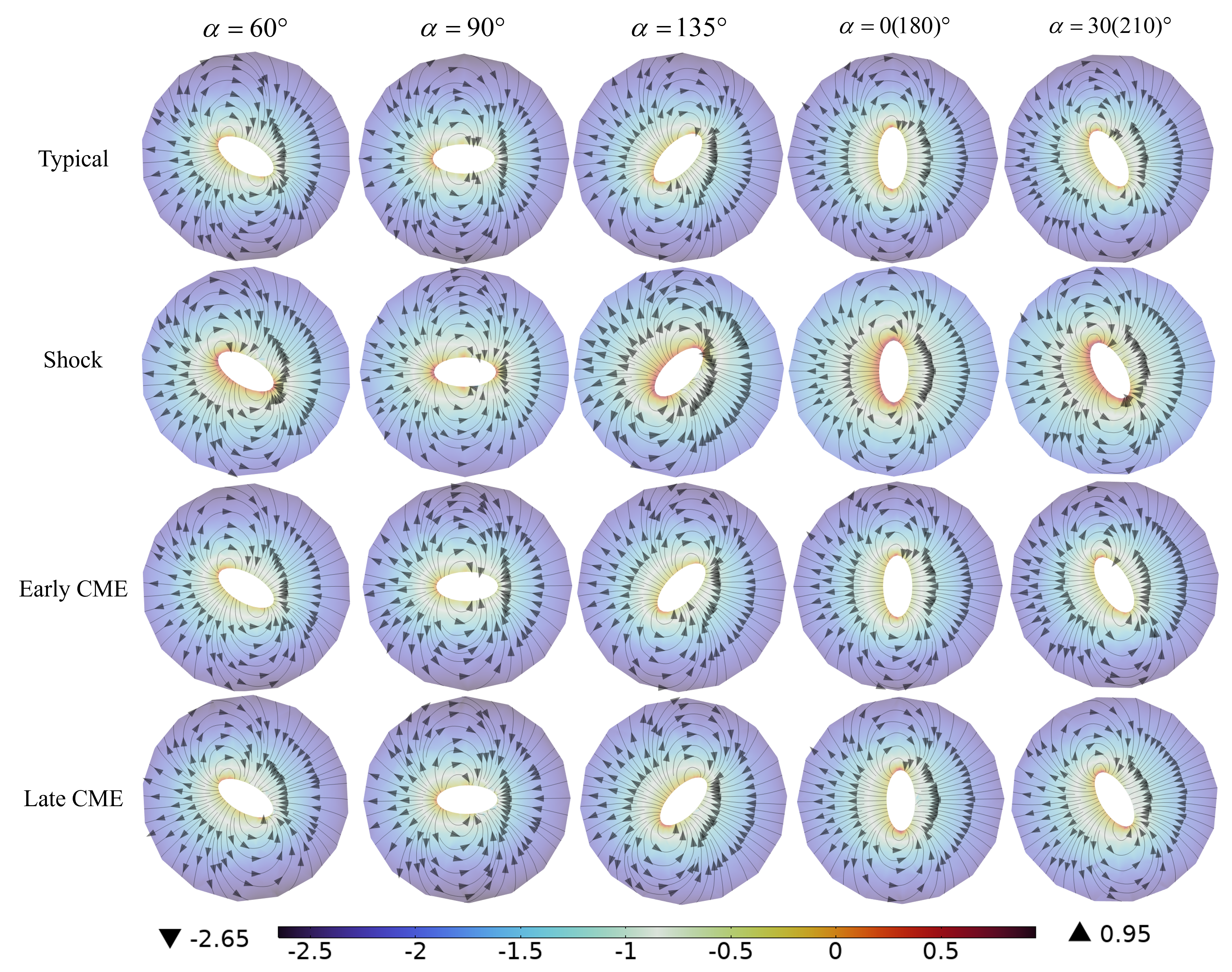}
\caption{Simulated electric field ($\log _{10}\left(E_{elef }\right)$) during passage of the modeled CME. The arrows indicate the direction of the electric field lines. We displayed the electric field in small-scale area (within 100m from the center of asteroid).
\label{fig:Fig9}}
\end{figure}

Figures \ref{fig:Fig8}-\ref{fig:Fig9} show results from a simulation with the occurrence of CME. Five attitudes are selected for display, in regard to typical solar wind, the dayside always charged positively, its potential depends on solar elevation angle, will decrease apparently during the process from subsolar point to the terminator, until it turns into a small negative potential on the nightside. When the forward shock exists, solar wind number density and temperature experience a sharp increase, solar wind starts to accelerate. At this point, a sudden drop can be seen in surface potential, photoelectron current is inferior to the current generated by high-energy electrons, the dayside is charged negatively, such as the potential of -52V when $\mathit{\alpha=60^{\circ}}$, the minimum potential on the nightside can reach -84.54V, close to 20 times that under normal circumstance, at this time, electron density here is $\mathit{\rm 2.91\times10^9/m^3}$. Therefore, even though at the same angle, potential difference between parts of asteroid's surface can exceed 40V, contributing to exponential growth of electric field strength, as seen in Figure \ref{fig:Fig9}, electric field strength near asteroid, particularly the terminator, where there is a sharp transition between positive and negative potentials, has reached the level of 10V/m, greatly enhancing the sensitivity of electrons and ions to the charged properties of asteroid, plasma is extremely unstable, thus the surface potential curve performs strong volatility. It only takes 1/12 period to make the average electron density increase from $\mathit{\rm 4.66\times10^8/m^3}$ to $\mathit{\rm 7.26\times10^8/m^3}$, bring about a potential drop of more than 10V.

\begin{figure}[ht!]
\centering
\includegraphics[width=0.7\linewidth]{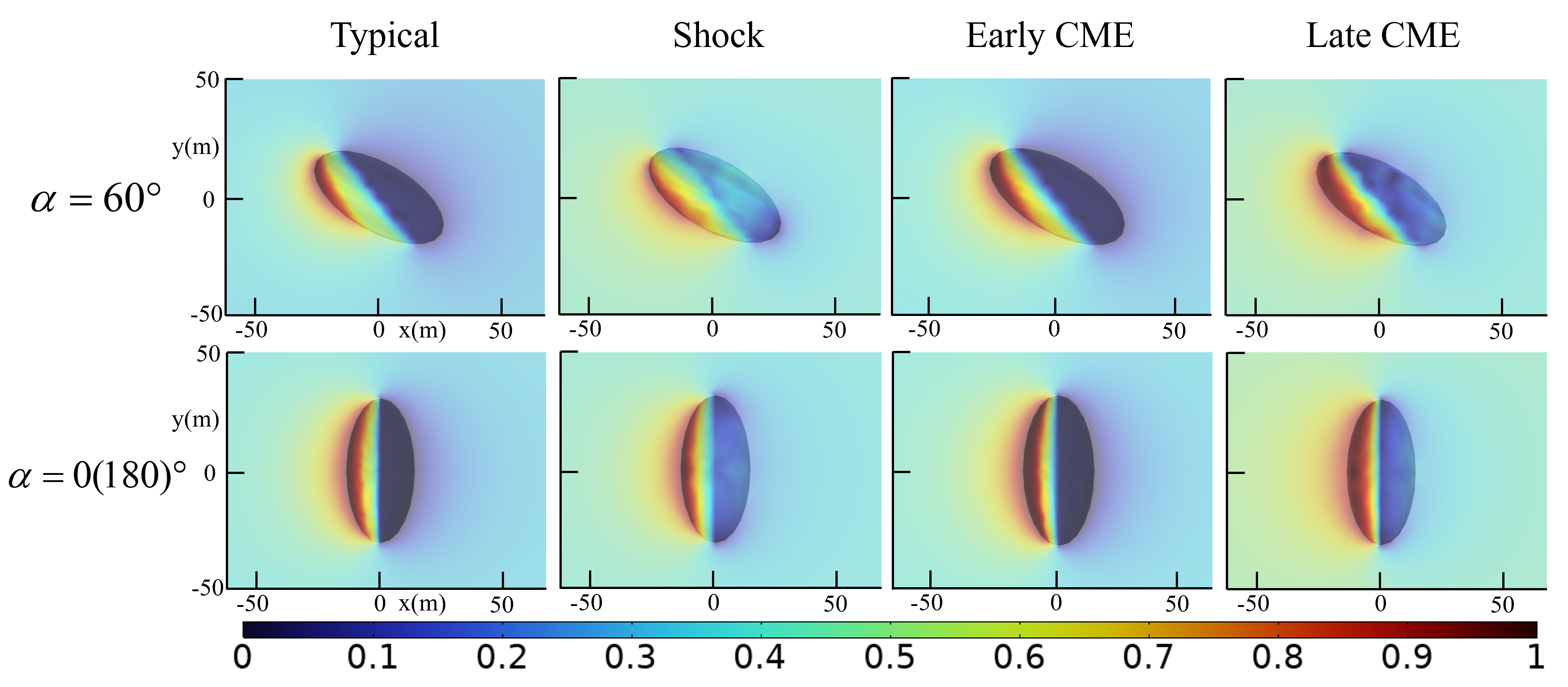}
\caption{Normalized potential during passage of the modeled CME. Potential around asteroid is scaled according to current $V_{\max}$ and $V_{\min}$, to more clearly demonstrate differences in distribution of surface potential under various environments.
\label{fig:Fig10}}
\end{figure}

In addition to magnitude, the distribution of potential is also worth investigating. Figure \ref{fig:Fig10} depicts the normalized potential, we can precisely analyze the potential distribution around asteroid, especially where the minimum potential is located. According to equation (\ref{eq:e11}-\ref{eq:e12}) and kinetic theory of plasma, both electrons and ions during shock achieve larger diffusion coefficients, the effect of drift-diffusion can partially overpower the flow of solar wind, increasing the density around asteroid's backside, then decrease surface potential there, allowing electrons to gather at the terminator, where has the strongest electric field. In contrast to the virtually uniform negative potential on the nightside during typical solar wind and early CME, potential during shock has a distinct gradient. In most cases, potential at the terminator is clearly lower than other parts, except $\mathit{\alpha=90^{\circ}}$, when solar elevation angle declines most slowly on the dayside, reducing the potential difference near the terminator, thereby weakening the electric field. Because of the fast-flowing solar wind, ions produce a considerable wake near asteroids tail, whereas high-temperature electrons do not. As a result, electron current far exceeds ion current, resulting in the lowest potential here.

The CME event has a lengthy duration, can be divided into early and late stages. Early CME exhibits characteristics of fast-flowing and cool, with a slightly lower number density than typical solar wind. Charging results at this stage have only minor differences from that of typical solar wind, in details, surface potential during early CME has a lift of 1$\sim$2V, mainly attributed to the low energy of charged particles. Photoelectron current on the dayside results in prominent positive potential. Furthermore, late CME produces more interesting phenomenon, the dense plasma flows slowly, surface of asteroid shrouded in it always has extreme negative potential, since electric field strength is not enhanced like that in shock, asteroid during late CME has more uniform surface potential. It is worth noting that surface potential will not reach its peak as in the former three stages, but it will continue to decrease. This is because for dense plasma, the equilibrium potential often needs to be several hundred volts, short-period asteroid is unable to reach this level within one period.

Then we change the rotation period. Charging results of asteroids with varying periods under typical solar wind have been discussed in \ref{subsec:SCP}. We now focus on the remaining three conditions. Apart from 0.467 hour, we select three additional periods for each stage based on their respective characteristics, results are shown in Table \ref{tab:Tab3}, including the maximum potential $V_{\max}$, the minimum potential $V_{\min}$, the maximum electric field strength $E_{elef\_\max}$, and the range of potential differences occur on asteroids.

\begin{table}[ht!]
\caption{Impact of asteroid's rotation on charging results during various solar storm stages}
\tabletypesize{\scriptsize}
\tablewidth{0pt}
    \centering
    \begin{tabular}{cccccc}
    \toprule
    Stage&	Period& $V_{\max}$(V)& $V_{\min}$(V)& $E_{elef\_\max}$(V/m)& $\Delta V$(V) \\
         \midrule
         \multirow{4}*{Shock}& 0.467 hour&	16.735&	-84.537&	12.623&	16.701$\sim$42.809  \\
         ~&1 hour&	16.731&	-98.132&	13.112&	16.393$\sim$42.340  \\
         ~&6 hours&	16.219&	-145.066&	14.234&	16.150$\sim$42.226  \\
         ~&1 day&	16.151&	-191.032&	12.722&	16.660$\sim$43.209  \\
         \midrule
\multirow{4}*{Early CME}& 0.467 hour& 9.568&	-3.729&	5.545&	12.094$\sim$12.906  \\
         ~&1 hour&	9.570&	-3.755&	5.494&	12.047$\sim$12.910  \\
         ~&12 hours&	9.570&	-3.755&	5.494&	12.111$\sim$12.910  \\
         ~&1 month&	9.568&	-3.763&	5.428&	12.110$\sim$12.908  \\
         \midrule
         \multirow{4}*{Late CME}& 0.467 hour& 5.615&	-73.799&	8.717&	5.196$\sim$18.890  \\
         ~&1 hour&	5.626&	-111.547&	10.638&	5.139$\sim$23.078  \\
         ~&2 hours&	5.413&	-179.863&	20.347&	5.026$\sim$34.108  \\
         ~&3 hours&	5.433&	-270.211&	18.802&	5.013$\sim$48.014  \\
         \bottomrule
    \end{tabular}
    \label{tab:Tab3}
\end{table}

Similar to typical solar wind, asteroids have diminishing $V_{\min}$ as their period grows, this trend can also be seen in $V_{\max}$, considering it always appears when surrounding plasma is thin, it won't develop to an uncommon value. At the stage shock, asteroid's surface retains a potential difference of over 40V, which does not change with the growth of rotation period, determined that there permanently exists strong electric field strength. Early CMEs are characterized by thin and fast-flowing, making it difficult for electrons and ions to collect near asteroid, the highest average electron density around asteroid is only $\mathit{\rm 1.10\times10^6/m^3}$ when period is 1 month. We are only able to identify a 0.008V reduction on the nightside when we obtain a large interval between periods. Conversely, a single hour of period expansion during late CME can lower $V_{\min}$ by 90.35V, together with the conspicuous increasing of $\Delta V$, therefore $E_{elef\_\max}$. It should be noted that for slow and dense solar wind, the rotation of asteroid will result in a significant change in surrounding environment, which was overshadowed by electric field migration when we study typical solar wind. At this stage, the minimum potential occurs at $\mathit{\alpha=0^{\circ}}$, i.e. $\mathit{\alpha=180^{\circ}}$, where asteroid has the largest direct contact area with solar wind, simultaneously owns the widest wake.

\subsection{Differences between Asteroids with Various Surface Minerals}
\label{subsec:DAV}

Except plagioclase, we studied orthopyroxene and ilmenite, obtain their charging results, then explore their sensitivity of surface potential to attitude of asteroids, their material parameters can be indicated in Table \ref{tab:Tab4}.

\begin{table}[ht!]
\caption{Material properties of different minerals}
\tabletypesize{\scriptsize}
\tablewidth{0pt}
    \centering
    \begin{tabular}{ccccc}
    \toprule
         Surface mineral&	Work function(eV)&	Threshold wavelength(nm)& $\mathit{\delta_{\text {max }}}$(eV)& $\mathit{E_{\text {max }}}$ \\
         \midrule
         Plagioclase&	5.58&	238&	2.8&	1000 \\
         Orthopyroxene&	5.14&	259&	2.1&	700 \\
         Ilmenite&	4.29&	297&	2.5&	600 \\
         \bottomrule
    \end{tabular}
    \label{tab:Tab4}
\end{table}

Solar wind parameters take that same as \ref{subsec:SCP}, rotation periods of asteroid are 1 hour, 6 hours and 1 day.

\begin{figure}[ht!]
\centering
\includegraphics[width=0.8\linewidth]{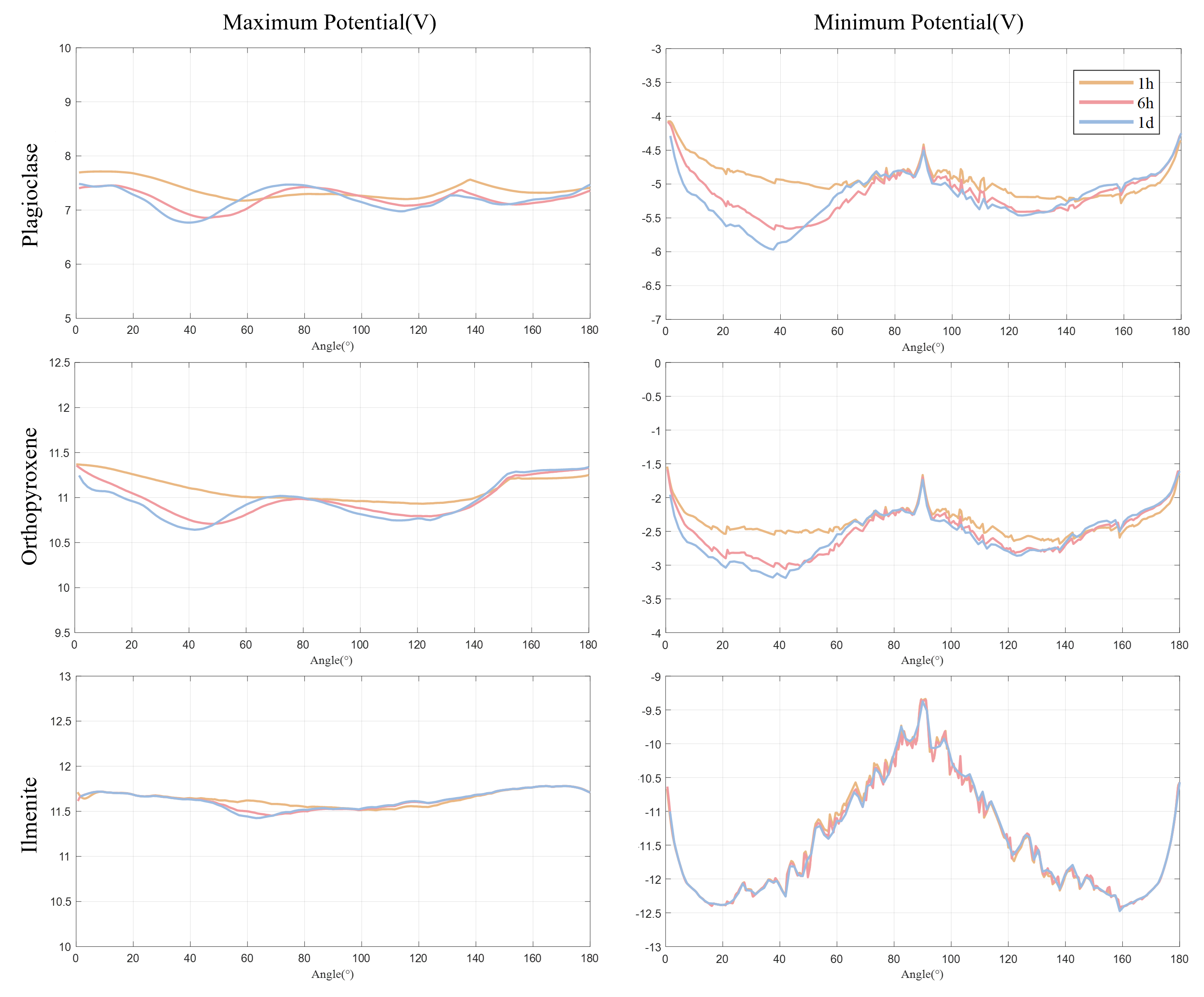}
\caption{Charging results of three constituent minerals of asteroids. On the left, there are curves showing the maximum surface potential of asteroids composed of three minerals. On the right, corresponding curves depict the minimum potential. All subfigures share a common legend.
\label{fig:Fig11}}
\end{figure}

Results are shown in Figure \ref{fig:Fig11}, we select rotation periods of 1 hour, 6 hours and 1 day as representations. Plagioclase is most susceptible to asteroid's rotation period among these minerals, while ilmenite seems indifferent to changes in rotation period, its three potential curves almost overlap. In this respect, orthopyroxene is analogous to plagioclase, as rotation period increases, long-period asteroid's potential is clearly lower.

About the dayside, we suppose that the intensity of solar radiation stays constant. Cause orthopyroxene and ilmenite have smaller work function than plagioclase, they gain stronger photoelectron currents, which results in a larger and more stable positive potential at the same time. This stability is improved from top to bottom in Figure \ref{fig:Fig11}, along with the decrease of work function.

Granted that there could be a potential difference of 3.1V on the nightside in the absence of CMEs, the impact of the asteroid's attitude on ilmenite is astounding. In contrast, potential of orthopyroxene changes gently, the greatest potential difference product by attitudes can only be 1.58V when rotation period is 1 day, even smaller than that of plagioclase.

\subsection{Effect of Orbital Motion and Obliquity on Surface Charging of Rotating Asteroids}
\label{subsec:EOM}

\begin{figure}[ht!]
\centering
\includegraphics[width=0.6\linewidth]{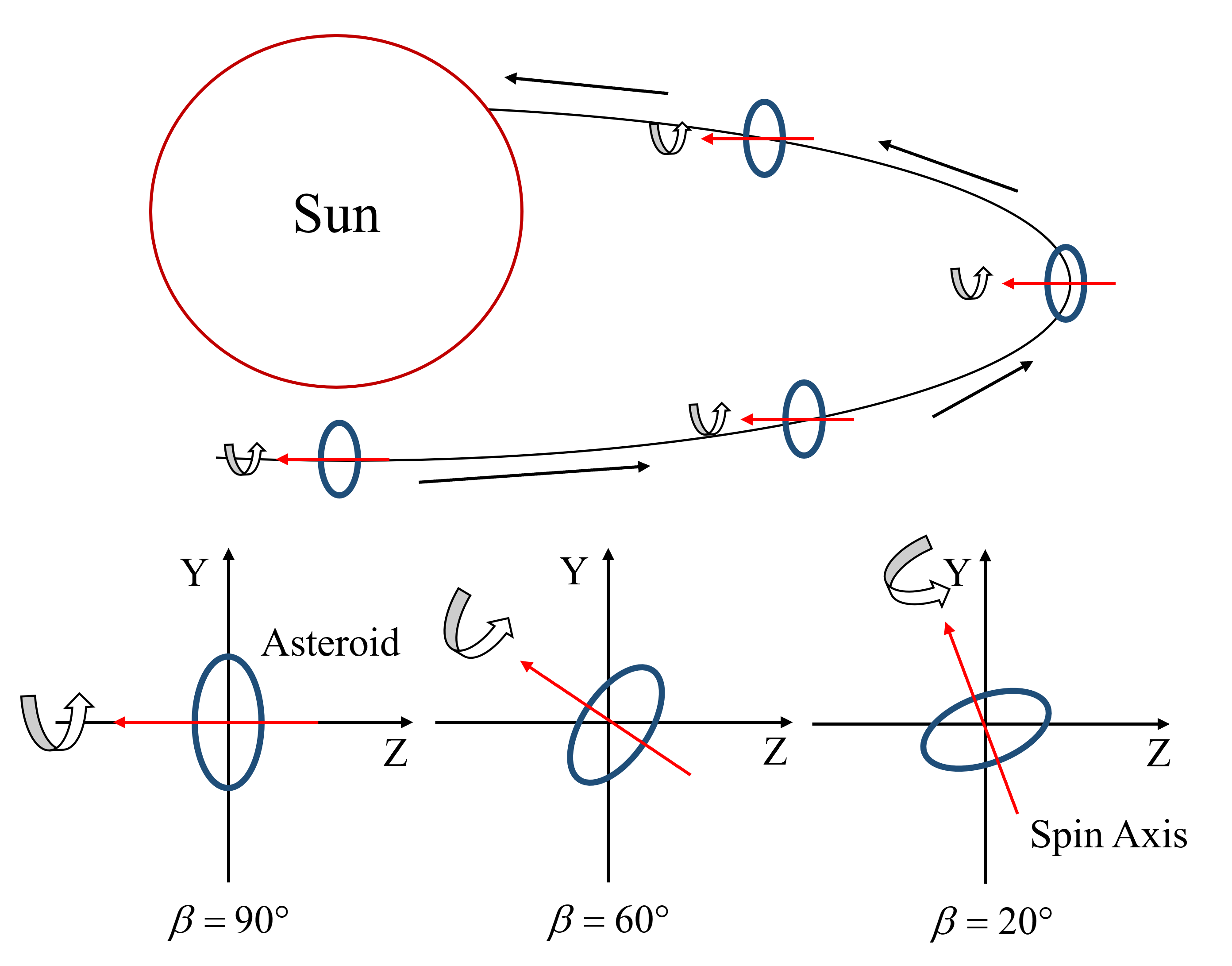}
\caption{Settings of orbital direction and spin axis obliquity. The blue ellipse represents asteroid, and the red arrow indicates the direction of spin axis.
\label{fig:Fig12}}
\end{figure}

\begin{figure}[ht!]
\centering
\includegraphics[width=0.8\linewidth]{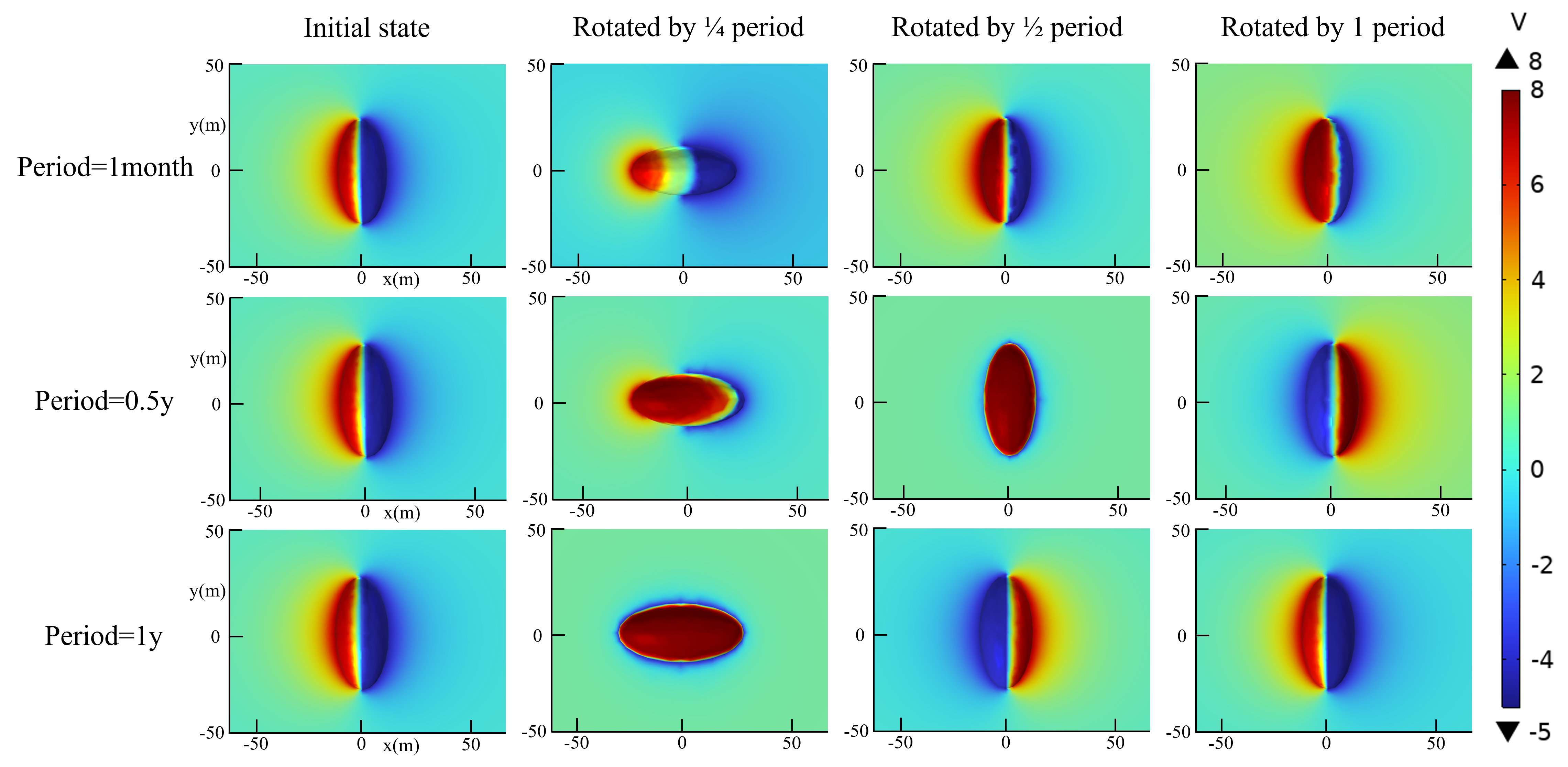}
\caption{Potentials of X-O-Y plane and asteroid's surface with three long rotation periods. Orbital period is 1 year, asteroid has a spin axis obliquity $\mathit{\beta=90^{\circ}}$, "Period" in the first column on the left represents asteroid's rotation period.
\label{fig:Fig13}}
\end{figure}

For asteroids with very long rotation periods, such as those longer than half a year, the variation in solar wind incidence angle due to orbital motion would become important. Therefore, we consider revolution based on the solar wind parameters in \ref{subsec:SCP}, plagioclase is selected as surface mineral of asteroid. Given that the orbital period of 2016HO3 is 365.9 days, we set one year as the orbital period of asteroid in the following studies, the direction of revolution is shown in Figure \ref{fig:Fig12}. We investigated surface charging of asteroids with rotation periods of 1 month, half a year, and 1 year under dynamic solar wind incidence directions. Results are shown in Figure \ref{fig:Fig13}.

Consistent with the conclusion in \ref{subsec:SCP}, the minimum potential that asteroid can reach during one rotation period still occurs when asteroid rotates to the attitude where the subsolar point is close to the terminator. For asteroid with period of 1 month, the impact of revolution is negligible. Compared with the situation where revolution is not considered in \ref{subsec:SCP}, the change in potential is within 0.05V. However, when asteroid has rotation period longer than half a year, the influence of revolution on surface charging is worth discussing. In addition to the effect on potential distribution in Figure \ref{fig:Fig13}, the minimum potential of asteroid's surface during simulation has also increased, from -5.96V in \ref{subsec:SCP} to -5.62V in \ref{subsec:EOM} for asteroid with period of half a year, from -5.96V to -5.53V for asteroid with period of 1 year, but remained well below the minimum potential that asteroids with short periods can achieve. Considering the constantly varying solar wind incidence angle, revolution can also be seen as a disruption and restructuring to the equilibrium state around asteroid, the motion of electrons and ions near asteroid is significantly disturbed, with velocity vectors varying much more dramatically than in \ref{subsec:SCP}. This mitigates the electron accumulation phenomenon around asteroid. Thus the potential curve becomes smoother. This effect is particularly noticeable when the rotation period of asteroid is comparable to the orbital period.

\begin{figure}[ht!]
\centering
\begin{subfigure}
{\includegraphics[width=0.75\linewidth]{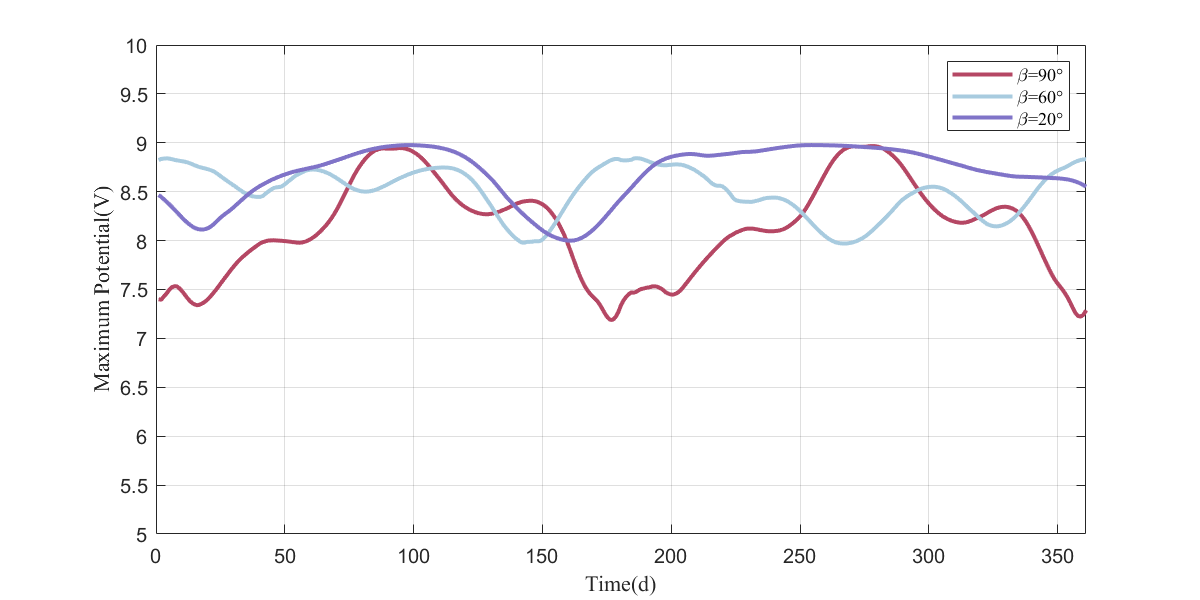}
}
\end{subfigure}

(a)

\begin{subfigure}
{\includegraphics[width=0.75\linewidth]{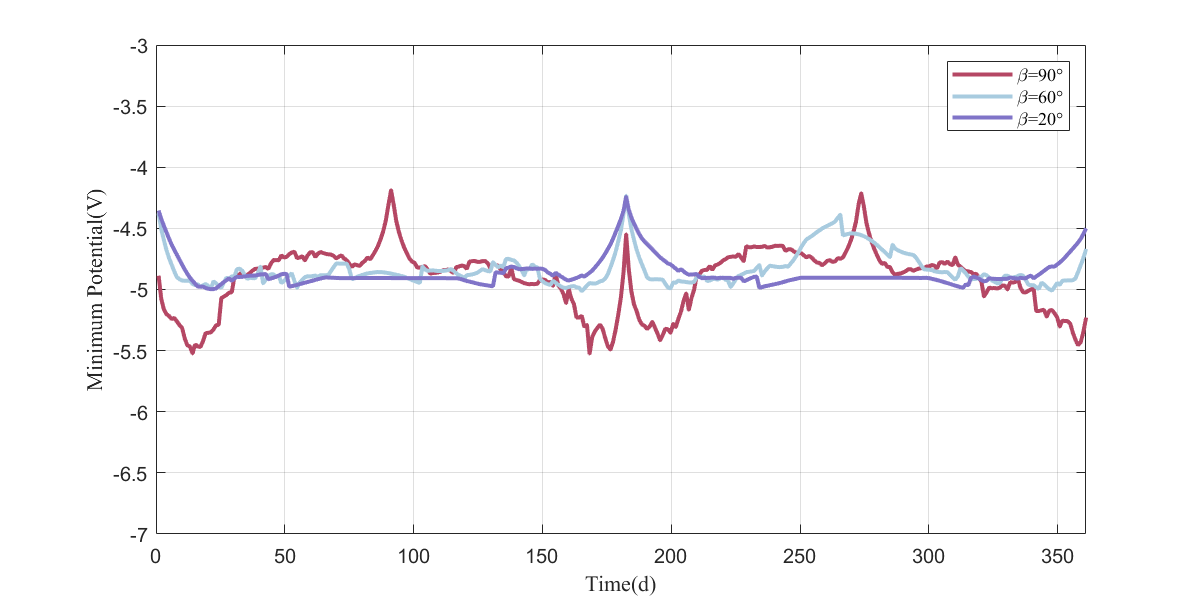}
}
\end{subfigure}

(b)

\caption{Potentials of asteroids with different spin axis obliquities. The maximum and minimum surface potentials of asteroids within one rotation period are shown. Their rotation period is 1 year.
\label{fig:Fig14}}
\end{figure}

Furthermore, in the above simulations, we assume that spin axis perpendicular to the X-O-Y plane, because this orientation allows asteroid to pass through some meaningful attitudes during rotation, such as those depicted in Figure \ref{fig:Fig8}, where $\mathit{\alpha=45^{\circ}}$ and $\mathit{\alpha=90^{\circ}}$. This orientation facilitates comparison and analysis of these special attitudes during asteroid's rotation. However, actually, asteroid's spin axis may have an inclination relative to the orbital plane. Consequently, as illustrated in Figure \ref{fig:Fig12}, we will alter the spin axis to explore surface potentials of asteroids under more common conditions with obliquities of $\mathit{\beta=90^{\circ}}$ (equivalent to previous simulation), $\mathit{\beta=60^{\circ}}$, and $\mathit{\beta=20^{\circ}}$. Results are shown in Figure \ref{fig:Fig14}.

If orbital motion is ignored, obliquity has little effect on the surface charging of asteroid. Therefore, we primarily discuss the surface potential of asteroids with different obliquities taking revolution into consideration. If spin axis obliquity $\mathit{\beta=90^{\circ}}$, asteroid's rotational motion and orbital motion occur in two mutually perpendicular planes, resulting in the greatest impact of revolution on asteroid's surface potential. As obliquity decreases, the potential curves gradually becomes gentler. When $\mathit{\beta=0^{\circ}}$, because both rotation period and orbital period of asteroid are one year, making the direction of solar wind incidence stationary relative to asteroid's surface. At this time, the maximum and minimum surface potentials of asteroid remain stable around 8.5V and -4.5V, respectively.

\section{Conclusion} \label{sec:Con}
We study how asteroid's rotation effect on its surface charging even surrounding plasma environment under various conditions. A multi-scale modeling method is proposed to implement dynamic three-dimensional simulation, we replace numerical calculation with neural network, improving the efficiency of our research. Rotation periods from 1 hour to half a year are all considered, it is found that rotation period of asteroid can affect its surface potential, specifically, strong electric field will accelerate electrons and ions, then result in higher density and temperature, the change in electrons is much more significant than that of ions, therefore reduce surface potential, until asteroid and plasma reach a dynamic balance, which will be continuously broken and reshaped after rotating. Both potentials on the dayside and nightside will decrease as asteroid's period increases, especially when asteroid rotates to the point where solar wind has an angle of $\mathit{\alpha=25\sim45^{\circ}}$ with it, the difference in period alone can bring a potential difference of 1V. This trend will gradually slow down with the increasing of period, eventually become stable when period exceeds 1 week, meaning that asteroids with these periods have the ability to reach balance at all angles. 

When the rotation period of asteroid is comparable to the orbital period, continuously varying solar wind incidence angle will act on the equilibrium state around asteroid, disturbing particles and making their motion more intense, thereby alleviating electron accumulation around the asteroid. Therefore, asteroids with periods longer than half a year experience a slight elevation in their surface potential. Furthermore, for asteroids with both rotation and orbital periods of 1 year, as spin axis obliquity decreases, the planes of rotation and orbital motion gradually align. Consequently, the potential curves exhibited on asteroid's surface within one period gradually become stable.

Solar wind parameters and material parameters are also closely related to surface charging. During the passage of solar storm, electron density and ion density become 200 and 10 times that under typical solar wind respectively, potential reached by an asteroid during its rotation can differ by more than 20V, product rapid changes in surrounding plasma, then react to surface potential. Dense and high-temperature plasma can charge asteroid to -84.54V with an electric field of 13V/m, while the potential is -4.78V in normal. As period increasing, asteroids at stage of shock and late CME behaves obvious fluctuations in potential. Different minerals can also exhibit vastly opposed charging results, both plagioclase and orthopyroxene more or less response to changes in rotation period, while ilmenite lacks this sensitivity, its potential mainly depends on the attitude of asteroids.

Our researches provide dynamic simulations of the interaction between solar wind plasma and electric field environments around asteroids, compensate for the lack of consideration in static simulations, namely rotation direction and angular velocity of asteroids. Studies about surface charging phenomenon of asteroids are important for future missions, have general implications in studying the charging properties of other small airless bodies.

\begin{acknowledgments}
This work was supported by the National Natural Science Foundation of China (grant No.42241148 and No.51877111).
\end{acknowledgments}

\bibliography{sample631}{}

\begin{thebibliography}{}
\expandafter\ifx\csname natexlab\endcsname\relax\def\natexlab#1{#1}\fi
\providecommand{\url}[1]{\href{#1}{#1}}
\providecommand{\dodoi}[1]{doi:~\href{http://doi.org/#1}{\nolinkurl{#1}}}
\providecommand{\doeprint}[1]{\href{http://ascl.net/#1}{\nolinkurl{http://ascl.net/#1}}}
\providecommand{\doarXiv}[1]{\href{https://arxiv.org/abs/#1}{\nolinkurl{https://arxiv.org/abs/#1}}}

\bibitem[{Adil {et~al.}(2022)Adil, Ullah, Noor, \& Gohar}]{adil2022effect}
Adil, M., Ullah, R., Noor, S., \& Gohar, N. 2022, Neural Computing and Applications, 34, 8355

\bibitem[{Eker {et~al.}(2023)Eker, Kayri, Ekinci, \& {\.I}zci}]{eker2023comparison}
Eker, E., Kayri, M., Ekinci, S., \& {\.I}zci, D. 2023, ADCAIJ: Advances in Distributed Computing and Artificial Intelligence Journal, 12, e29969

\bibitem[{{Farrell} {et~al.}(2012){Farrell}, {Halekas}, {Killen}, {Delory}, {Gross}, {Bleacher}, {Krauss-Varben}, {Travnicek}, {Hurley}, {Stubbs}, {Zimmerman}, \& {Jackson}}]{2012JGRE..117.0K04F}
{Farrell}, W.~M., {Halekas}, J.~S., {Killen}, R.~M., {et~al.} 2012, Journal of Geophysical Research (Planets), 117, E00K04, \dodoi{10.1029/2012JE004070}

\bibitem[{{Ginzburg}(1970)}]{1970pewp.book.....G}
{Ginzburg}, V.~L. 1970, {The propagation of electromagnetic waves in plasmas}

\bibitem[{{Gorin} {et~al.}(2020){Gorin}, {Kudryavtsev}, {Yao}, {Yuan}, \& {Zhou}}]{2020PhPl...27a3505G}
{Gorin}, V.~V., {Kudryavtsev}, A.~A., {Yao}, J., {Yuan}, C., \& {Zhou}, Z. 2020, Physics of Plasmas, 27, 013505, \dodoi{10.1063/1.5120613}

\bibitem[{{Gurevich}(1978)}]{1978npi..book.....G}
{Gurevich}, A.~V. 1978, {Nonlinear phenomena in the ionosphere}, Vol.~10, \dodoi{10.1007/978-3-642-87649-3}

\bibitem[{Halekas {et~al.}(2009)Halekas, Delory, Lin, Stubbs, \& Farrell}]{halekas2009lunar}
Halekas, J., Delory, G., Lin, R., Stubbs, T., \& Farrell, W. 2009, Journal of Geophysical Research: Space Physics, 114

\bibitem[{{Halekas} {et~al.}(2007){Halekas}, {Delory}, {Brain}, {Lin}, {Fillingim}, {Lee}, {Mewaldt}, {Stubbs}, {Farrell}, \& {Hudson}}]{2007GeoRL..34.2111H}
{Halekas}, J.~S., {Delory}, G.~T., {Brain}, D.~A., {et~al.} 2007, \grl, 34, L02111, \dodoi{10.1029/2006GL028517}

\bibitem[{Hartzell \& Scheeres(2013)}]{hartzell2013dynamics}
Hartzell, C.~M., \& Scheeres, D.~J. 2013, Journal of Geophysical Research: Planets, 118, 116

\bibitem[{{Hastings} \& {Garrett}(2004)}]{2004sei..book.....H}
{Hastings}, D., \& {Garrett}, H. 2004, {Spacecraft-Environment Interactions}

\bibitem[{Kili{\c{c}}arslan(2023)}]{kiliccarslan2023novel}
Kili{\c{c}}arslan, S. 2023, Multimedia Tools and Applications, 82, 6345

\bibitem[{{Kureshi} {et~al.}(2020){Kureshi}, {Tripathi}, \& {Mishra}}]{2020Ap&SS.365...23K}
{Kureshi}, R., {Tripathi}, K.~R., \& {Mishra}, S.~K. 2020, \apss, 365, 23, \dodoi{10.1007/s10509-020-3740-8}

\bibitem[{Li \& Scheeres(2021)}]{LI2021114249}
Li, X., \& Scheeres, D.~J. 2021, Icarus, 357, 114249, \dodoi{https://doi.org/10.1016/j.icarus.2020.114249}

\bibitem[{{Liu} {et~al.}(2022){Liu}, {Xu}, {Wang}, {Ding}, \& {Xiao}}]{2022MRE.....9b5504L}
{Liu}, H., {Xu}, Y., {Wang}, C., {Ding}, F., \& {Xiao}, H. 2022, Materials Research Express, 9, 025504, \dodoi{10.1088/2053-1591/ac3a40}

\bibitem[{{Novikov} {et~al.}(2008){Novikov}, {Mileev}, {Krupnikov}, {Makletsov}, {Marjin}, {Rjazantseva}, {Sinolits}, \& {Vlasova}}]{2008AdSpR..42.1307N}
{Novikov}, L.~S., {Mileev}, V.~N., {Krupnikov}, K.~K., {et~al.} 2008, Advances in Space Research, 42, 1307, \dodoi{10.1016/j.asr.2008.02.019}

\bibitem[{{Oudayer} {et~al.}(2019){Oudayer}, {Monnin}, {Mateo-Velez}, {Hess}, {Sarrailh}, {Murat}, \& {Roussel}}]{2019ITPS...47.3710O}
{Oudayer}, P., {Monnin}, L., {Mateo-Velez}, J.~C., {et~al.} 2019, IEEE Transactions on Plasma Science, 47, 3710, \dodoi{10.1109/TPS.2019.2919932}

\bibitem[{Pandya {et~al.}(2019)Pandya, Mehta, \& Kothari}]{Impact}
Pandya, A., Mehta, P., \& Kothari, N. 2019, International Journal of Numerical Modelling: Electronic Networks, Devices and Fields, 32, e2631, \dodoi{https://doi.org/10.1002/jnm.2631}

\bibitem[{Qian {et~al.}(2023)Qian, Zhang, Huang, Huang, \& Wang}]{qian2023accuracy}
Qian, H.-M., Zhang, H., Huang, T., Huang, H.-Z., \& Wang, K. 2023, Quality and Reliability Engineering International, 39, 1878

\bibitem[{{Quan} {et~al.}(2023){Quan}, {Zhang}, \& {Zhang}}]{2023ITPS...51.1181Q}
{Quan}, R., {Zhang}, C., \& {Zhang}, H. 2023, IEEE Transactions on Plasma Science, 51, 1181, \dodoi{10.1109/TPS.2023.3254516}

\bibitem[{{Skoug} {et~al.}(1999){Skoug}, {Bame}, {Feldman}, {Gosling}, {McComas}, {Steinberg}, {Tokar}, {Riley}, {Burlaga}, {Ness}, \& {Smith}}]{1999GeoRL..26..161S}
{Skoug}, R.~M., {Bame}, S.~J., {Feldman}, W.~C., {et~al.} 1999, \grl, 26, 161, \dodoi{10.1029/1998GL900207}

\bibitem[{{Stubbs} {et~al.}(2014){Stubbs}, {Farrell}, {Halekas}, {Burchill}, {Collier}, {Zimmerman}, {Vondrak}, {Delory}, \& {Pfaff}}]{2014P&SS...90...10S}
{Stubbs}, T.~J., {Farrell}, W.~M., {Halekas}, J.~S., {et~al.} 2014, \planss, 90, 10, \dodoi{10.1016/j.pss.2013.07.008}

\bibitem[{Wang {et~al.}(2016)Wang, Wu, Tang, Yi, \& Sun}]{7407629}
Wang, S., Wu, Z.-C., Tang, X.-J., Yi, Z., \& Sun, Y.-W. 2016, IEEE Transactions on Plasma Science, 44, 289, \dodoi{10.1109/TPS.2016.2521867}

\bibitem[{{Whipple}(1981)}]{1981RPPh...44.1197W}
{Whipple}, E.~C. 1981, Reports on Progress in Physics, 44, 1197, \dodoi{10.1088/0034-4885/44/11/002}

\bibitem[{{Xie} {et~al.}(2023){Xie}, {Li}, {Wang}, {Zhang}, {Zhou}, \& {Feng}}]{2023ApJ...952...61X}
{Xie}, L., {Li}, L., {Wang}, J., {et~al.} 2023, \apj, 952, 61, \dodoi{10.3847/1538-4357/acd6ec}

\bibitem[{Xin-Yue {et~al.}(2016)Xin-Yue, Ai-Bing, Tao, Reme, Ling-Gao, Shen-Yi, \& Chun-Lai}]{Synchronization}
Xin-Yue, W., Ai-Bing, Z., Tao, J., {et~al.} 2016, Chinese Journal of Geophysics (in Chinese), 59, 3533, \dodoi{10.6038/cjg20161001}

\bibitem[{Zhu {et~al.}(2023)Zhu, Cui, Liu, Jiang, Liu, \& Wang}]{zhu2023method}
Zhu, H., Cui, Z., Liu, J., {et~al.} 2023, Journal of Marine Science and Engineering, 11, 1340

\bibitem[{{Zimmerman} {et~al.}(2016){Zimmerman}, {Farrell}, {Hartzell}, {Wang}, {Horanyi}, {Hurley}, \& {Hibbitts}}]{2016JGRE..121.2150Z}
{Zimmerman}, M.~I., {Farrell}, W.~M., {Hartzell}, C.~M., {et~al.} 2016, Journal of Geophysical Research (Planets), 121, 2150, \dodoi{10.1002/2016JE005049}

\bibitem[{{Zimmerman} {et~al.}(2014){Zimmerman}, {Farrell}, \& {Poppe}}]{2014Icar..238...77Z}
{Zimmerman}, M.~I., {Farrell}, W.~M., \& {Poppe}, A.~R. 2014, \icarus, 238, 77, \dodoi{10.1016/j.icarus.2014.02.029}

\bibitem[{{Zimmerman} {et~al.}(2012){Zimmerman}, {Jackson}, {Farrell}, \& {Stubbs}}]{2012JGRE..117.0K03Z}
{Zimmerman}, M.~I., {Jackson}, T.~L., {Farrell}, W.~M., \& {Stubbs}, T.~J. 2012, Journal of Geophysical Research (Planets), 117, E00K03, \dodoi{10.1029/2012JE004094}

\end{thebibliography}
\bibliographystyle{aasjournal}

\end{document}